\documentclass[12pt]{article}
\usepackage[affil-it]{authblk}
\usepackage[colon]{natbib}
\usepackage{tabularx}
\usepackage{epsfig}
\usepackage[titletoc, title]{appendix}
\usepackage{setspace}
\usepackage{lscape}
\usepackage{booktabs}
\usepackage{array}
\usepackage{subfig}
\usepackage{tabularx}
\usepackage[latin1]{inputenc}
\usepackage[english]{babel}
\usepackage{makeidx}
\usepackage{latexsym}
\usepackage{ifthen}
\usepackage{color,epsfig}
\usepackage{graphicx}
\usepackage{mathrsfs}
\usepackage{wrapfig}
\usepackage{rotating}
\usepackage{natbib}
\usepackage{amsmath, amsthm}
\usepackage{amssymb}
\usepackage{nomencl}
\usepackage{float}
\usepackage{multirow}
\usepackage{multicol}
\usepackage{xcolor,soul}
\usepackage{epic,eepic}
\usepackage{lscape}
\usepackage{tabularx}
\usepackage{pdflscape}
\usepackage{epstopdf}
\usepackage{calligra}
\usepackage{epsfig}
\usepackage[colon]{natbib}
\usepackage{tabularx}
\usepackage{amsmath}
\usepackage{amssymb}
\usepackage{pdflscape}
\usepackage{multirow}
\usepackage{multicol}
\usepackage[bottom]{footmisc}
\usepackage{array,arydshln}
\usepackage{appendix}
\usepackage[plain,noend]{algorithm2e}

\SetAlCapSkip{1em}

\newcommand{\beq}{\begin{equation}}
\newcommand{\eeq}{\end{equation}}
\newcommand{\bed}{\begin{displaymath}}
\newcommand{\eed}{\end{displaymath}}
\begin{document}

\newcommand{\bs}{\boldsymbol } 
\newcommand{\tbf}{\textbf } 

\newcommand{\U}{{\cal U}}
\newcommand{\R}{{\cal R}}
\newcommand{\Sc}{{\cal S}}

\newcommand{\abs}[1]{\left\vert#1\right\vert}
\newcommand{\set}[1]{\left\{#1\right\}}
\newcommand{\bra}[1]{\left(#1\right)}
\newcommand{\braq}[1]{\left[#1\right]}

\newcommand{\eps}{\varepsilon}
\newcommand{\si}{\sigma_q}
\newcommand{\ka}{\kappa}
\newcommand{\tr}{^{T}}
\newcommand{\bbeta}{\boldsymbol{\beta}}
\newcommand{\blambda}{\boldsymbol{\lambda}}
\newcommand{\bpi}{\boldsymbol{\pi}}
\newcommand{\bphi}{\boldsymbol{\phi}}
\newcommand{\bpsi}{\boldsymbol{\psi}}
\newcommand{\bxi}{\boldsymbol{\xi}}
\newcommand{\bptheta}{\boldsymbol \theta}
\newcommand{\bomega}{\boldsymbol\omega}
\newcommand{\bDelta}{\boldsymbol\Delta}
\newcommand{\bPhi}{\boldsymbol\Phi}
\newcommand{\bPsi}{\boldsymbol\Psi}
\newcommand{\boldeta}{\boldsymbol\eta}
\newcommand{\bgamma}{\boldsymbol{\gamma}}
\newcommand{\bzeta}{\boldsymbol{\zeta}}
\newcommand{\beps}{\boldsymbol{\varepsilon}}
\newcommand{\bSigma}{\boldsymbol{\Sigma}}
\newcommand{\bUpsilon}{\boldsymbol{\Upsilon}}

\def \cI {{\cal I}}

\def \bA {{\mathbf A}}
\def \bB {{\mathbf B}}
\def \bC {{\mathbf C}}
\def \bD {{\mathbf D}}
\def \bb {{\mathbf b}}
\def \be {{\mathbf e}}
\def \b
f {{\mathbf f}}
\def \bFF {{\mathbf F}}
\def \bg {{\mathbf g}}
\def \bG {{\mathbf G}}
\def \bH {{\mathbf H}}
\def \bh {{\mathbf h}}
\def \bI {{\mathbf I}}
\def \bL {{\mathbf L}}
\def \br {{\mathbf r}}
\def \bP {{\mathbf P}}
\def \bQ {{\mathbf Q}}
\def \bt {{\mathbf t}}
\def \bX {{\mathbf X}}
\def \bx {{\mathbf x}}
\def \by {{\mathbf y}}
\def \bu {{\mathbf u}}
\def \bv {{\mathbf v}}
\def \bV {{\mathbf V}}
\def \bZ {{\mathbf Z}}
\def \bz {{\mathbf z}}
\def \bU {{\mathbf U}}
\def \bW {{\mathbf W}}
\def \bw {{\mathbf w}}
\def \bzero {{\mathbf 0}}

\def \diag {\mbox{diag}}
\def \uno {\mathbf 1}
\def\b#1{\mbox{\boldmath $#1$}}    
\def\bl#1{\mbox{\footnotesize \boldmath {$#1$}}} 
\def\m#1{\mbox{#1}}                
\def\ml#1{\mbox{\scriptsize #1}} 
\def\ha#1{\mbox{$\hat{\b #1}$}}
\def \bPsi {\boldsymbol\Psi}

\newcommand{\sumuni}{\sum_{i\in\U}}
\newcommand{\sums}{\sum_{i\in s}}
\newcommand{\emmqr}{$\mathrm{FMMQ_R}$}
\newcommand{\emmqe}{$\mathrm{FMMQ_E}$}
\newcommand{\emmq}{$\mathrm{FMMQ}$}
\newcommand{\qrre}{\textsf{QRRE}}
\newcommand{\qr}{\textsf{QR}}
\newcommand{\pqr}{\textsf{PQR}}
\newcommand{\mq}{\textsf{MQ}}
\newcommand{\lmm}{\textsf{LRE}}
\newcommand{\elmm}{\textsf{ERE}}
\newcommand{\btau}{\b{\tau}}
\newcommand{\tcr}{\textcolor{red}}
\newcommand{\tcb}{\textcolor{blue}}
\newcommand{\tcg}{\textcolor{green}}
\definecolor{green}{rgb}{0., 0.5, 0}

\title{A bi-dimensional finite mixture model for longitudinal \\data subject to  dropout}
\author{Alessandra Spagnoli}
\affil{Dipartimento di Sanit\`a Pubblica e Malattie Infettive, Sapienza Universit\`a di Roma,  \texttt{alessandra.spagnoli@uniroma1.it}}
\author{Maria Francesca Marino}
\affil{Dipartimento di Statistica, Informatica, Applicazioni, Universit\`a  degli Studi di Firenze, \texttt{mariafrancesca.marino@unifi.it}}
\author{Marco Alf\`o }
\affil{Dipartimento di Scienze Statistiche, Sapienza Universit\`a di Roma, \texttt{marco.alfo@uniroma1.it}}

\date{}
\maketitle

\begin{abstract}
In longitudinal studies, subjects may be lost to follow-up, or miss some of the planned visits,  leading to incomplete response sequences. When the probability of non-response, conditional on the available covariates and the observed responses, still depends on unobserved outcomes, the dropout mechanism is said to be \emph{non ignorable}. A common objective is to build a reliable association structure to account for dependence between the longitudinal and the dropout processes. Starting from the existing literature, we introduce a random coefficient based dropout model where the association between outcomes is modeled through discrete latent effects. These effects are outcome-specific and account for heterogeneity in the univariate profiles. Dependence between profiles is introduced by using a bi-dimensional representation for the corresponding distribution. In this way, we define a flexible latent class structure which allows to efficiently describe both dependence \textit{within} the two margins of interest and dependence \textit{between} them. 
By using this representation we show that, unlike standard (unidimensional) finite mixture models, the non ignorable dropout model properly nests its ignorable counterpart. We detail the proposed modeling approach by analyzing data from a longitudinal study on the dynamics of cognitive functioning in the elderly. Further, the effects of assumptions about non ignorability of the dropout process on model parameter estimates are (locally) investigated using the index of (local) sensitivity to non-ignorability.
\end{abstract}


\section{Introduction}
In longitudinal studies, measurements from the same individuals (units) are repeatedly taken over time. However, individuals may be lost to follow up or do not show up at some of the planned measurement occasions, leading to attrition (also referred to as \emph{dropout}) and intermittent missingness, respectively. \citet{rub1976} provides a well-known taxonomy for mechanisms that generate incomplete sequences. If the probability of a missing response does not depend neither on the observed nor on the missing responses, conditional on the observed covariates, the data are said to be missing completely at random (MCAR). Data are missing at random (MAR) if, conditional on the observed data (both covariates and responses),  the missingness does not depend on the non-observed responses. When the previous assumptions do not hold, that is when, conditional on the observed data, the mechanism leading to missing data still depends on the unobserved responses, data are referred to as missing not at random (MNAR). In the context of likelihood inference, when the parameters in the measurement and in the missingness processes are distinct, processes leading either to MCAR or MAR data may be ignored; when either the parameter spaces are not distinct or the missing data process is MNAR, missing data are non-ignorable (NI). Only when the ignorability property is satisfied, standard (likelihood) methods can be used to obtain consistent parameter estimates. Otherwise, some form of joint modeling of the longitudinal measurements and the  missigness process is required. See \citet{litrub2002} for a comprehensive review of the topic. 

For this purpose, in the following, we will focus on the class of Random Coefficient Based Dropout Models \citep[RCBDMs - ][]{Little1995}. 
In this framework, separate (conditional) models are built for the two partially observed processes, and the link between them is due to sharing common or dependent individual- (and possibly outcome-) specific random coefficients. The model structure is completed by assuming that the random coefficients are drawn from a given probability distribution. 
Obviously, a choice is needed to define such a distribution and, in the past years, the literature focused both on parametric and nonparametric specifications. Frequently, the random coefficients are assumed to be Gaussian \citep[e.g.][]{ver2002, gao2004}, but this assumption was questioned by several authors, see e.g. \citet{sch1999}, since the resulting inference can be sensitive to such assumptions, especially in the case of short longitudinal sequences. For this reason, \citet{alf2009} proposed to leave the random coefficient distribution unspecified, defining a semi-parametric model where the longitudinal and the dropout processes are linked through dependent (discrete) random coefficients. \cite{tso2009} suggested to follow a similar approach for handling intermittent, potentially non ignorable, missing data. 
A similar approach to deal with longitudinal Gaussian data subject to missingness was proposed by \citet{Beunc2008}, where a finite mixture of mixed effect regression models for the longitudinal and the dropout processes was discussed.  
Further generalizations in the shared parameter model framework were proposed by \citet{cre2011}, who discussed an approach based on \emph{partially} shared individual (and outcome) specific random coefficients, and by \citet{bart2015} who extended standard latent Markov models to handle potentially informative dropout, via shared discrete random coefficients.

In the present paper, the association structure between the measurement and the dropout processes is based on a random coefficient distribution which is left completely unspecified, and estimated through a discrete distribution, leading to a (bi-dimensional) finite mixture model. The adopted bi-dimensional structure allows the bivariate distribution for the random coefficients to reduce to the product of the corresponding marginals when the dropout mechanism is ignorable. Therefore, a peculiar feature of the proposed modeling approach, when compared to standard finite mixture models, is that the MNAR specification properly nests the MAR/MCAR ones, and this allows a straightforward (local) sensitivity analysis. We propose to explore the sensitivity of parameter estimates in the longitudinal model to the assumptions on non-ignorability of the dropout process by developing an appropriate version of the so-called \emph{index of sensitivity to non-ignorability} (ISNI) developed by \cite{trox2004} and \cite{ma2005}, considering different perturbation scenarios.

{The structure of the paper follows. In section \ref{sec:2} we introduce the motivating application, the Leiden 85+ study, entailing the dynamics of cognitive functioning in the elderly. Section \ref{sec:3} discusses general random coefficient based dropout models, while our proposal is detailed in section\ref{sec:4}. Sections \ref{sec:5}-\ref{sec:6} detail the proposed EM algorithm for maximum likelihood estimation of model parameters and the index of local sensitivity we propose. Section \ref{sec:7} provides the application of the proposed model to data from the motivating example, using either MAR or MNAR assumptions, and the results from sensitivity analysis. Last section contains concluding remarks.
}

\section{Motivating example: Leiden 85+ data}
\label{sec:2}
The motivating data come from the Leiden 85+ study, a retrospective study entailing 705 Leiden inhabitants (in the Netherlands), who reached the age of 85 years between September 1997 and September 1999. The study aimed at identifying demographic and genetic determinants for the dynamics of cognitive functioning in the elderly. Several covariates collected at the beginning of the study were considered: gender {(female is the reference category)}, educational status distinguishing between primary {(reference category)} or higher education, plasma Apolipoprotein E (APOE) genotype. As regards the educational level, this was determined by the number of years each subject went to school; primary education corresponds to less than 7 years of schooling. As regards the APOE genotype, the three largest groups were considered: $\epsilon2,\epsilon3$, and $\epsilon 4$. This latter allele is known to be linked to an increased risk for dementia, whereas $\epsilon 2$ allele carriers are relatively protected. Only 541 subjects present complete covariate information and will be considered in the following. 

Study participants were visited yearly until the age of $90$ at their place of residence and face-to-face interviews were conducted through a questionnaire whose items are designed to assess orientation, attention, language skills and the ability to perform simple actions. The Mini Mental State Examination index, in the following MMSE \citep{fol1975}, is obtained by summing the scores on the items of the questionnaire designed to assess potential cognitive impairment. The observed values are integers ranging between $0$ and $30$ (maximum total score). 

\begin{figure}[h]
    \centering
    \subfloat[]{{\includegraphics[width=6.5cm]{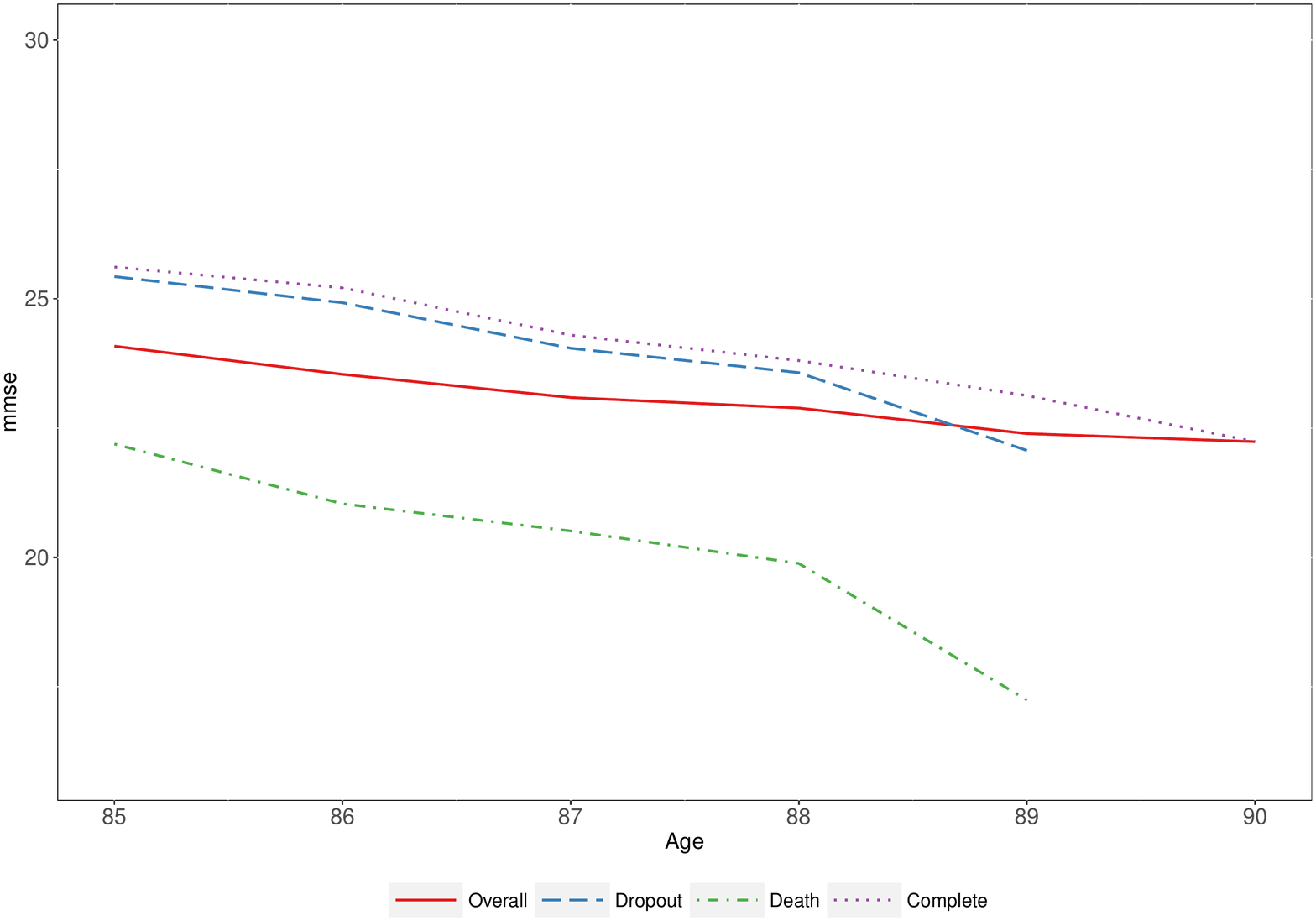} }}%
    \quad
    \subfloat[]{{\includegraphics[width=6.5cm]{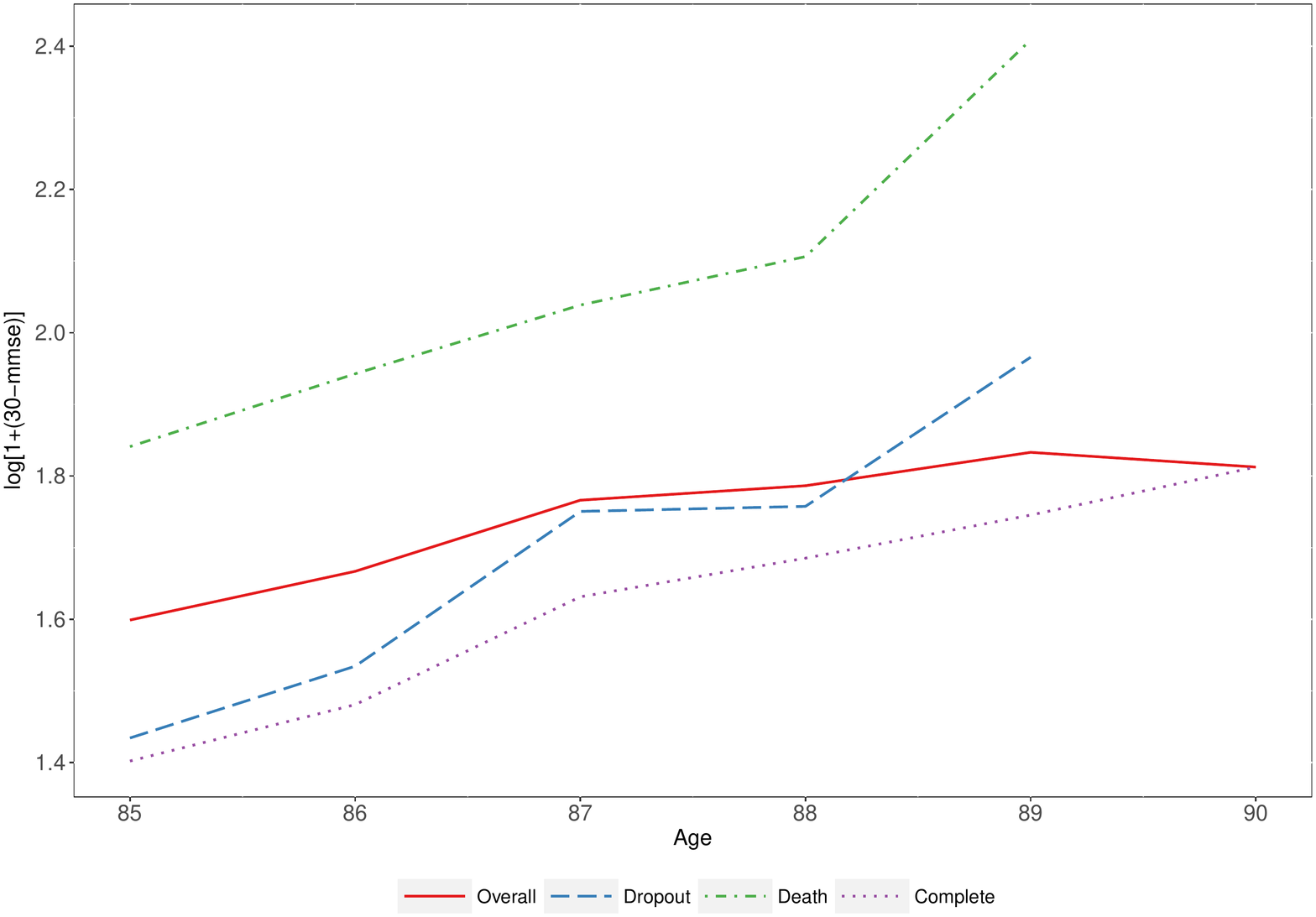} }}%
\caption{Mean MMSE (a) and mean $\log[1+(30-MMSE)]$ (b) over time stratified by subjects' participation to the study.}
\label{fig:mean}
\centering
\end{figure}

A number of enrolled subjects dropout prematurely, because of poor health conditions or death. In Table \ref{tab1}, we report the total number of available measures for each follow-up visit. Also, we report the number (and the percentage) of participants who leave the study between the current and the subsequent occasion, distinguishing between those who dropout and those who die. 
As it can be seen, less than half of the study participants presents complete longitudinal sequences ($49\%$) and this is mainly due to death ($44\%$ of the subjects died during the follow-up). 
\begin{table}[h]
\begin{center}
\caption{Available measures per follow-up visit and number (percentage) of subjects leaving the study between subsequent occasions due to poor health conditions or death}
\label{tab1}
\vspace{1mm}
\begin{tabular}{l c c c c c } \hline 
Follow-up      & Total & Complete (\%)  &  Do not  (\%)  &  Die (\%) \\
age               &      &              &       participate                   &                \\ \hline
85-86             & 541   & 484 (89.46) &        9 (1.66)       &    48 (8.87)  \\
86-87             & 484   & 422 (87.19) &        3 (0.62)       &    59 (12.19) \\
87-88             & 422   & 373 (88.39) &        2 (0.47)       &    47 (11.14)\\
88-89             & 373   & 318 (85.25) &        6 (1.61)       &    49  (13.14) \\
89-90             & 318   & 266 (83.65) &       15 (4.72)       &    37 (11.63) \\
 \hline
Total             & 541   & 266 (0.49) &        35 (0.07)       &   240 (0.44) \\
 \hline 
\end{tabular}
\end{center}
\end{table}
With the aim at understanding how the MMSE score evolves over time, we show in Figure \ref{fig:mean} the corresponding overall mean value across follow-up visits. We also represent the evolution of the mean MMSE stratified by participation in the study (completer, dropout/death before next occasion). As it is clear, cognitive functioning levels in individuals who die are much lower than those corresponding to subjects who dropout for other reasons or participate until the study end. The same figure is represented by considering the transform $\log[1+(30-MMSE)]$ which is negatively proportional to the MMSE score but will be further considered as it avoids well known ceiling and floor effects that are usually faced when dealing with this kind of indexes.

A further empirical evidence which is worth to be observed is that, while the decline in cognitive functioning (as measured by the MMSE score) through time seems to be (at least approximately) constant across groups defined by patterns of dropout, the differential participation in the study leads to a different slope when the overall mean score is considered. Such a finding highlights a potential dependence between the evolution of the MMSE score over time and the dropout process, which may bias parameter estimates and corresponding inference. In the next sections, we will introduce a bi-dimensional finite mixture model for the analysis of longitudinal data subject to potentially non-ignorable dropouts.

\section{Random coefficient-based dropout models}
\label{sec:3}
Let $Y_{it}$ represent a set of longitudinal measures recorded on $i=1,\ldots,n,$ subjects at time occasions $t=1,\ldots,T$,  and let $\mathbf{x}_{it}=(x_{it1},\ldots,x_{itp})^{\prime}$ denote the corresponding $p$-dimensional vector of observed covariates. Let us assume that, conditional on a $q$-dimensional set of individual-specific random coefficients ${\bf b}_{i}$, the observed responses $y_{it}$ are independent realizations of a random variable with density in the Exponential Family. The canonical parameter $\theta_{it}$ that indexes the density is specified according to the following regression model:
\begin{equation*}
\theta_{it}=\mathbf{x}_{it}^{\prime} \boldsymbol{\beta} + \mathbf{m}_{it}^{\prime}\mathbf{b}_i.
\label{modlong}
\end{equation*}
The terms $\mathbf{b}_i$, $i=1,\ldots,n,$ are used to model the effects of unobserved individual-specific, time-invariant, heterogeneity common to each lower-level unit (measurement occasion) within the same $i$-th upper-level unit (individual). Furthermore, $\boldsymbol{\beta}$ is a $p-$dimensional vector of regression parameters that are assumed to be constant across individuals. The covariates whose effects (are assumed to) vary across individuals are collected in the design vector $\mathbf{m}_{it}=(m_{it1},\ldots,m_{itq})$, which represents a proper/improper subset of $\mathbf{x}_{it}$. For identifiability purposes, standard assumptions on the random coefficient vector are 
$$
\textrm{E}({\bf b}_{i})={\bf 0}, \quad \textrm{Cov}({\bf b}_{i})={\bf D}, \quad i=1,\dots,n.
$$
Experience in longitudinal data modeling suggests that a potential major issue when dealing with  such a kind of studies is represented by missing data. That is, some individuals enrolled in the study do not reach its end and, therefore, only partially participate in the planned sequence of measurement occasions. 
In this framework, let $\mathbf{R}_i$ denote the missing data indicator vector, with generic element $R_{it}=1$ if the $i$-th unit drops-out at any point in the window $(t-1,t)$, and $R_{it}=0$ otherwise. As we remarked above, we consider a discrete time structure for the study and the time to dropout; however, the following arguments may apply, with a limited number of changes, to continuous time survival process as well. We assume that, once a person drops out, he/she is out forever (attrition). Therefore, if the designed completion time is denoted by $T$, we have, for each participant, $T_i\leq T$ available measures. 

To describe the (potential) dependence between the primary (longitudinal) and the secondary (dropout) processes, we may introduce an explicit  model for the dropout mechanism, conditional on a set of dropout specific covariates, say $\mathbf{v}_i$, and (a subset of) the random coefficients in the longitudinal response model:
\begin{equation}
h(\mathbf{r}_i \mid \mathbf{v}_i,\mathbf{y}_i,\mathbf{b}^{\ast}_i)  =  h(\mathbf{r}_i \mid \mathbf{v}_i,\mathbf{b}^{\ast}_i)= \prod_{t=1}^{\min(T, T_i+1)} h(r_{it} \mid \mathbf{v}_i,\mathbf{b}^{\ast}_i), \qquad i=1,\ldots,n. 
\label{drop}
\end{equation}
The distribution is indexed by a canonical parameter defined via the regression model:
\[
\phi_{it}=\mathbf{v}_{it}^\prime\boldsymbol{\gamma}+\mathbf{d}_{it}^\prime\mathbf{b}^{\ast}_i
\]
where ${\bf b}_{i}^{\ast}={\bf C} {\bf b}_{i}$, $i=1,\dots,n$, and $\bf C$ is a binary $q_{1}$-dimensional matrix ($q_1 \leq q$), with at most a one in each row.
These models are usually referred to as shared (random) coefficient models; see \cite{wu1988}, \cite{wu1989} for early developments in the field. As it may be evinced from equation \eqref{drop}, the assumption of this class of models is that the longitudinal response and the dropout indicator are independent conditional on the individual-specific random coefficients. According to this (local independence) assumption, the joint density of the observed longitudinal responses and and the missingness indicator can be specified as
\begin{eqnarray*}
f({\bf y}_{i}, {\bf r}_{i} \mid \mathbf{X}_{i},\mathbf{V}_i)&=&\int f({\bf y}_{i}, {\bf r}_{i} \mid \mathbf{X}_{i},\mathbf{V}_i, \mathbf b_i) dG(\mathbf{b}_i) = \nonumber \\
&=& \int \left[ \prod_{t=1}^{T_{i}} f(y_{it} \mid \mathbf{x}_{it},\mathbf{b}_i) \prod_{t=1}^{\min(T, T_i+1)} h(r_{it} \mid \mathbf{v}_{it}, \mathbf{b}_i) \right]dG(\mathbf{b}_i),
\label{joint}
\end{eqnarray*} 
where $G(\cdot)$ represents the random coefficient distribution, often referred to as the \emph{mixing} distribution. Dependence between the measurement and the missigness, if any, is completely accounted for by the latent effects which are also used to describe unobserved, individual-specific, heterogeneity in each of the two (univariate) profiles. 

As it can be easily noticed, this modeling structure leads to a perfect correlation between (subsets of) the random coefficients in the two equations, and this may not be a general enough setting. As an alternative, we may consider equation-specific random coefficients. In this context, while the random terms describe univariate heterogeneity and overdispersion, the corresponding joint distribution allows to model the association between the random coefficients in the two equations and, therefore, between the longitudinal and  the missing data process (on the link function scale). \cite{ait2003} discussed such an alternative parameterization referring to it as the \emph{correlated} random effect model. To avoid any confusion with the estimator proposed by \cite{cham1984}, we will refer to it as the \emph{dependent} random coefficient model. When compared to \emph{shared} random coefficient  models, this approach avoids unit correlation between the random terms in the two equations and, therefore, it represents a more flexible approach, albeit still not general. 

Let $\mathbf{b}_i=(\mathbf{b}_{i1},\mathbf{b}_{i2})$ denote a set of individual- and outcome-specific random coefficients. Based on the standard local independence assumption, the joint density for the couple $\left({\bf Y}_{i}, {\bf R}_{i}\right)$ can be factorized as follows:
\begin{equation}
f({\bf y}_{i}, {\bf r}_{i} \mid \mathbf{X}_{i},\mathbf{V}_i)=\int  \left[ \prod_{t=1}^{T_{i}} f(y_{it}|\mathbf{x}_{it},\mathbf{b}_{i1})\prod_{t=1}^{\min(T, T_i+1)} h(r_{it}|\mathbf{v}_{it},\mathbf{b}_{i2}) \right] dG(\mathbf{b}_{i1},\mathbf{b}_{i2}).
\label{joint_corr}
\end{equation}
A different approach to dependent random coefficient models may be defined according to the general scheme proposed by \cite{cre2011}, where common, partially shared and independent (outcome-specific) random coefficients are considered in the measurement and the dropout process. This approach leads to a particular case of dependent random coefficients where, however, the observed and the missing part of the longitudinal response do not generally come from the same distribution.

\subsection{The random coefficient distribution}
When dealing with dependent random coefficient models, a common assumption is that outcome-specific random coefficients are iid Gaussian variates. 
According to \cite{wan2001}, \cite{son2002}, \cite{tsi2004}, \cite{nehu2011a}, \cite{nehu2011b} the choice of the random effect distribution may not have great impact on parameter estimates, except in extreme cases, e.g. when the \emph{true} underlying distribution is discrete. In this perspective, a major role is played by the individual sequence length: when all subjects have  a relatively large number of repeated measurements, the effects of misspecifying the random effect distribution on model parameter estimates becomes minimal; see the discussion in \cite{riz2008}, who designed a simulation study to investigate the effects that a misspecification of the random coefficient distribution may have on parameter estimates and corresponding standard errors when a shared parameter model is considered. The authors showed that, as the number of repeated measurements per individual grows, the effect of misspecifying the random coefficient  distribution vanishes for certain parameter estimates. These results are motivated by making explicit reference to theoretical results in \citet{car2000}. In several contexts, however, the follow-up times may be short (e.g. in clinical studies) and individual sequences include only limited information on the random coefficients. In these cases, assumptions on such a distribution may play a crucial role. As noticed by \cite{tso2009}, the choice of an \emph{appropriate} distribution is generally difficult for, at least, three reasons; see also \citet{alf2009}. First, there is often little information about unobservables in the data, and any assumption is difficult to be justified by looking at the observed data only. Second, when high dimensional random coefficients are considered, the use of a parametric multivariate distribution imposing the same shape on every dimension can be restrictive. Last, a potential dependence of the random coefficients on omitted covariates induces heterogeneity that may be hardly captured by parametric assumptions. In studies where subjects have few measurements, the choice of the random coefficient distribution may therefore be extremely important. 

With the aim at proposing a generally applicable approach, \cite{tso2009} considered a semi-parametric approach with shared (random) parameters to analyze continuous longitudinal responses while adjusting for non monotone missingness. On the same line, \cite{alf2009} discussed a model for longitudinal binary responses subject to dropout, where dependence is described via outcome-specific, dependent, random coefficients. 
According to these finite mixture-based approaches and starting from equation \eqref{joint_corr}, we may write the observed data log-likelihood function as follows:
\begin{align}
\ell(\bPsi, \bPhi, \bpi) & = \sum_{i=1}^n \log \left\lbrace \sum_{k=1}^K f(\mathbf{y}_i \mid \mathbf{X}_i,\bzeta_{1k})
h(\mathbf{r}_i \mid \mathbf{V}_i,\bzeta_{2k})\pi_k \right\rbrace 
\nonumber \\
& = \sum_{i=1}^n \log \left\lbrace \sum_{k=1}^K \left[\prod_{t = 1}^{T_i} f({y}_{it} \mid \mathbf{X}_i,\bzeta_{1k})
\prod_{t=1}^{\min(T, T_i +1)} h({r}_{it} \mid \mathbf{V}_i,\bzeta_{2k})\pi_k \right]\right\rbrace,
\label{eq:log-likelihood}
\end{align}
where $\b \Psi=(\bbeta, \bzeta_{11}, \dots,\bzeta_{1K})$, $\b \Psi=(\bgamma, \bzeta_{21}, \dots,\bzeta_{2K})$, with $\bzeta_{1k}$ and $\bzeta_{2k}$ denoting the vectors of discrete random coefficients in the longitudinal and in the missingness process, respectively. 
Last, $\bpi=(\pi_{1},\dots,\pi_{K})$, with $\pi_k=\Pr({\bf b}_{i}=\bzeta_k)=\Pr\left({\bf b}_{i1}=\bzeta_{1k},{\bf b}_{i2}=\bzeta_{2k}\right)$ identifies the joint probability of (multivariate) locations $\bzeta_k=\left(\bzeta_{1k},\bzeta_{2k}\right)$, $k=1,\ldots,K$.
It is worth noticing that, in the equation above, ${\bf y}_{i}$ refers to the observed individual sequence, say $\mathbf{y}_i^{o}$. Under the model assumptions and due to the local independence between responses coming from the same sample unit, missing data, say $\mathbf{y}_i^{m}$, can be directly integrated out from the joint density of all longitudinal responses, say $f(\mathbf{y}_i^{o}, \mathbf{y}_i^{m} \mid \mathbf{x}_i, \bzeta_{k})$, and this leads to the log-likelihood function in equation \eqref{eq:log-likelihood}. 

The use of finite mixtures has several significant advantages over parametric approaches. Among others, the EM algorithm for ML estimation is computationally efficient and the discrete nature of the estimates may help classify subjects into disjoint components that may be interpreted as clusters of individuals characterized by homogeneous values of model parameters. However, as we may notice by looking at expression in equation (\ref{eq:log-likelihood}), the latent variables used to account for individual (outcome-specific) departures from homogeneity are intrinsically uni-dimensional. That is, while the locations may differ across profiles, the number of locations ($K$) and the prior probabilities ($\pi_k$) are common to all profiles. This clearly reflects the standard \emph{unidimensionality} assumption in latent class models. 

\section{A bi-dimensional finite mixture approach}
\label{sec:4}
Although the  modeling approach described above is quite general and flexible, a clear drawback is related to the non-separability of the association structure  between the random coefficients in the longitudinal and the missing data profiles. 
Moreover, even if it can be easily shown that the likelihood in equation (\ref{eq:log-likelihood}) refers to a MNAR model in Rubin's taxonomy \citep{litrub2002}, it is also quite clear that it does not reduce to the (corresponding) MAR model, but in very particular cases (e.g. when $K=1$ or when either $\bzeta_{1k}=cost$ or $\bzeta_{2k}=cost$, $\forall k=1,\dots,K$). This makes the analysis of sensitivity to modeling assumptions complex to be exploited and, therefore, makes the application scope of such a modeling approach somewhat narrow.

Based on the considerations above and with the aim at enhancing model flexibility, we suggest to follow an approach similar to that proposed by \cite{alf2012}. That is, we consider outcome-specific sets of discrete random coefficients for the longitudinal and the missingness outcome, where each margin is characterized by a (possibly) different number of components. Components in each margin are successively joined by a full (bi-dimensional) matrix containing the masses associated to couples $(g,l)$, where the first index refers to the components in the longitudinal response profile, while the second denotes the components in the dropout process. 

To introduce our proposal, let $\tbf b_i = (\tbf b_{i1}, \tbf b_{i2})$ denote the vector of individual random coefficients associated to the $i$-th subject, with $i=1, \dots n$. 
Let us assume the vector of individual-specific random coefficients $\tbf b_{1i}$ influences the longitudinal data process and follows a discrete distribution defined on $K_1$ distinct support points $\{\b \zeta_{11}, \dots, \b \zeta_{1K_1}\}$ with masses $\pi_{g\star} = \Pr(\tbf b_{i1} = \b\zeta_{1g})$. 
Similarly, let us assume the vector of random coefficients $\tbf b_{2i}$ influences the missing data process and follows a discrete distribution with $K_2$ distinct support points $\{\b \zeta_{21}, \dots, \b \zeta_{1K_2}\}$ with masses $\pi_{\star l} = \Pr(\tbf b_{i2} = \b\zeta_{2l})$. 
That is, we assume that
\[
\tbf b_{1i} \sim \sum_{g = 1}^{K_1} \pi_{g\star} \, \delta(\b \zeta_{1g}), \quad \quad \tbf b_{2i} \sim \sum_{\ell = 1}^{K_2} \pi_{\star \ell} \, \delta(\b \zeta_{2\ell}),
\]
where $\delta(a)$ denotes an indicator function that puts unit mass at $a$. 

To complete the modeling approach we propose, we introduce a joint distribution for the random coefficients, associating a mass $\pi_{g\ell}=\Pr({\bf b}_{i1}=\bzeta_{1g}, {\bf b}_{i2}=\bzeta_{2\ell})$ to each couple of locations $(\bzeta_{1g}, \bzeta_{2\ell})$, entailing the longitudinal response and the dropout process, respectively.
Obviously, masses $\pi_{g\star}$ and $\pi_{\star \ell}$ in the univariate profiles are obtained by properly marginalizing $\pi_{g\ell}$: 
\[
\pi_{g\star} = \sum_{\ell = 1}^{K_2} \pi_{g\ell}, \quad \quad 
\pi_{\star \ell} = \sum_{g = 1}^{K_1} \pi_{g\ell}.
\]

Under the proposed model specification, the likelihood in equation (\ref{eq:log-likelihood}) can be written as follows:
\begin{equation}
\ell(\bPhi, \bPsi, \bpi) = \sum_{i=1}^n \log \left\lbrace \sum_{g=1}^{K_1} \sum_{\ell=1}^{K_2} \left[f(\mathbf{y}_i \mid \mathbf{x}_i,\bzeta_{1g})h(\mathbf{r}_i \mid \mathbf{v}_i,\bzeta_{2l})\right] \pi_{g\ell}  \right\rbrace.
\label{eq:log-likelihooddouble}
\end{equation}
Using this approach, marginals control for heterogeneity in the univariate profiles, while the matrix of joint probabilities $\pi_{gl}$ describes the association between the latent effects in the two sub-models. 
The proposed specification could be considered as a standard finite mixture with $K = K_{1} \times K_{2}$ components, where each of the $K_{1}$ locations in the first profile pairs with each of the $K_{2}$ locations in the second profile. 
However, when compared to a standard finite mixture model, the proposed specification provides a more flexible (albeit more complex) representation for the random coefficient distribution. Also, by looking at equation \eqref{eq:log-likelihooddouble}, it is immediately clear that the MNAR model directly reduces to its M(C)AR counterpart when $\pi_{g\ell}=\pi_{g\star} \pi_{\star \ell}$, for $g=1,\dots,K_{1}$ and $\ell=1,\dots,K_{2}$. As we stressed before, this is not true in the case of equation \eqref{eq:log-likelihood}.

Considering a logit transform for the joint masses $\pi_{g\ell}$, we may write
\begin{equation}\label{pi:equation}
\xi_{g \ell}=\log \left(\frac{\pi_{g \ell}}{\pi_{K_{1} K_{2}}}\right) = \alpha_{g \star} + \alpha_{\star \ell} + \lambda_{g \ell},
\end{equation}
where
$\alpha_{g \star} = \log \left({\pi_{g \star}}/{\pi_{K_{1} \star}}\right), \alpha_{\star \ell} = \log \left({\pi_{\star \ell}}/{\pi_{\star K_{2}}}\right),$
and $\lambda_{g \ell}$ provides a measures of the departure from independence model. 
That is, if $\lambda_{g \ell}=0$, for all $(g, \ell) \in \left\{1,\dots,K_{1}\right\} \times \left\{1,\dots,K_{2}\right\}$, then 
\[
\log \left(\frac{\pi_{g \ell}}{\pi_{K_{1} K_{2}}}\right) = \alpha_{g \star} + \alpha_{\star \ell}=\log \left(\frac{\pi_{g \star}}{\pi_{K_{1} \star}}\right)+\log \left(\frac{\pi_{\star \ell}}{\pi_{\star K_{2}}}\right).
\]

This corresponds to the independence between the random coefficients in the two equations, and, as a by-product, to the independence between the longitudinal and the dropout process. 
Therefore, the vector $\blambda = (\lambda_{11}, \dots, \lambda_{K_1K_2})$ can be formally considered as a \textit{sensitivity} parameter vector, since when $\blambda=\b 0$ the proposed MNAR model reduces to the corresponding M(C)AR model.

It is worth noticing that the proposed approach has some connections with the model discussed by \citet{Beunc2008}, where parametric shared random coefficients for the longitudinal response and the dropout indicator are joined by means of a (second-level) finite mixture. In fact, according to Theorem 1 in \cite{dun2009}, the elements of any $K_{1} \times K_{2}$ probability matrix $\mbox{\boldmath$\Pi$} \in \mathcal{M}_{K_{1}K_{2}}$, can be decomposed as:
\begin{equation}
\pi_{g\ell}=\sum_{h=1}^{M} \tau_{h} \pi_{g \star \mid h} \pi_{\star \ell \mid h},
\label{eq:pidecomp}
\end{equation}
for an appropriate choice of $M$ and under the following constraints: \begin{eqnarray*}
\sum_{h} \tau_{h}=\sum_{g} \pi_{g\star \mid h}=\sum_{\ell} \pi_{\star \ell \mid
h}=\sum_{g} \sum_{\ell} \pi_{g\ell}=1.
\end{eqnarray*}

If we use the above parameterization for the random coefficient distribution, the association between locations $\bzeta_{1g}$ and $\bzeta_{2\ell}$, $g=1,\dots,K_{1}$ $\ell=1,\dots,K_{2}$ is modeled via the masses $\pi_{g \star \mid h}$ and $\pi_{\star \ell \mid h}$, that vary according to the upper-level (latent) class $h=1,\dots,M$. 
That is, random coefficients $\mathbf{b}_{1i}$ and $\mathbf{b}_{2i}$, $i=1,\dots,n$ are assumed to be independent conditional on the $h$-th (upper level) latent class $h=1,\dots,M$. Also, the mean and covariance matrix in profile-specific random coefficient distribution may vary with second-level component, while in the approach we propose, the second level structure is just a particular way to join the two profiles and, therefore, control for dependence between outcome-specific random coefficients.

\section{ML parameter estimation}
\label{sec:5}
Let us start by assuming that the data vector is composed by an observable part $(\mathbf{y}_{i}, \mathbf{r}_i)$ and by unobservables $\mathbf{z}_{i}=(z_{i11},\dots,z_{ig\ell}, \dots, z_{iK_1K_2})$.  Let us further assume the random variable $\mathbf{z}_{i}$ has a multinomial distribution, with parameters $\pi_{g\ell}$ denoting the probability of the $g$-th component in the first and the $\ell$-th component in the second profile, for $g=1, \dots, K_1$ and $\ell = 1, \dots, K_2$,. 
Let $\bUpsilon = \left\{\bPhi, \bPsi, \bpi\right\}$ denote the vector of all (free) model parameters, where, as before, $\boldsymbol{\Phi} = (\bbeta, \bzeta_{11}, \dots, \bzeta_{1K_1})$ and $\boldsymbol \Psi = (\bgamma, \bzeta_{21}, \dots, \bzeta_{2K_2})$ collect the parameters for the longitudinal and the missing data model, respectively, and $\b \pi= (\pi_{11}, \dots, \pi_{K_1 K_2})$.
Based on the modeling assumptions introduced so far, the complete data likelihood function is given by
\begin{eqnarray*}
L_{c}(\bUpsilon)& = & \prod_{i=1}^{n}\prod_{g=1}^{K_{1}}\prod_{\ell=1}^{K_{2}}\left\{f(\mathbf{
y}_{i},\mathbf{r}_i\mid z_{ig\ell} = 1) \pi_{g\ell}\right\}^{z_{ig\ell}} 
\\
& =& \prod_{i=1}^{n}\prod_{g=1}^{K_{1}}\prod_{\ell=1}^{K_{2}}\left\{ 
\left[ \prod_{t=1}^{T_{i}} f(y_{it}\mid z_{ig\ell}=1 )
\prod_{t=1}^{\min(T, T_{i}+1)}  h(r_{it}\mid z_{ig\ell}=1)\right]\pi_{g\ell} \right\}^{z_{ig\ell}}.
\end{eqnarray*}

To derive parameter estimates, we can exploit an extended EM algorithm which, as usual, alternates two separate steps. In the ($r$-th iteration of the) E-step, we compute the posterior expectation of the complete data log-likelihood, conditional on the observed data $(\mathbf{y}_{i}, \mathbf{r}_i)$ and the current parameter estimates $\hat{\boldsymbol{\bUpsilon}}^{(r-1)}$. This translates into the computation of the posterior probability of component membership $w_{ig\ell}$, defined as posterior expectation of the random variable $z_{ig\ell}$.
In the M-step, we maximize the expected complete-data log-likelihood with respect to model parameters. Clearly, for the finite mixture probabilities $\pi_{g\ell}$, estimation is based upon the constraint $\sum_{g=1}^{K_1}\sum_{\ell=1}^{K_2} \pi_{g\ell} = 1$. As a result, the following score functions are obtained: 
\begin{eqnarray*}
\label{eq:score1}
S_{c}(\boldsymbol{\Phi})&=& \sum\limits_{i=1}^{n} \frac{\partial }{\partial {\boldsymbol \Phi}} \sum\limits_{g=1}^{K_{1}} \sum\limits_{\ell=1}^{K_{2}} w_{ig\ell}^{(r)} \left[\log(f_{ig\ell}) + \log(\pi_{g\ell})
\right]=
\sum\limits_{i=1}^{n} \frac{\partial }{\partial {\boldsymbol \Phi}} \sum\limits_{g=1}^{K_{1}}  w_{ig\star}^{(r)}\left[\log(f_{i1g})\right], \\ 
\label{eq:score2}
S_{c}(\boldsymbol{\Psi})&=&\sum\limits_{i=1}^{n}\frac{\partial }{\partial {\boldsymbol \Psi}}  \sum\limits_{g=1}^{K_{1}}\sum\limits_{\ell=1}^{K_{2}} w_{ig\ell}^{(r)}\left[\log(f_{ig\ell}) + \log(\pi_{g\ell})
\right]=
\sum\limits_{i=1}^{n} \frac{\partial }{\partial {\boldsymbol \Psi}}\sum\limits_{\ell=1}^{K_{2}} w_{i\star \ell}^{(r)} \left[\log(f_{i2\ell})\right], \\
\label{eq:score3}
S_{c}({\pi}_{g\ell})&=&\sum\limits_{i=1}^{n} \frac{\partial }{\partial {\pi}_{g\ell}} \sum_{g=1}^{K_{1}} \sum_{\ell=1}^{K_{2}} w_{ig\ell}^{(r)} \pi_{g\ell} - \kappa \left(
\sum_{g=1}^{K_{1}} \sum_{\ell=1}^{K_{2}} \pi_{g\ell} -1
 \right).
\end{eqnarray*}
In the equations above, $f_{ig\ell} = f(\textbf y_{i}, \textbf r_i \mid z_{ig\ell})$, while $w_{ig\star}^{(r)}$, $w_{i\star l}^{(r)}$, $f_{i1g}$ and $f_{i2\ell}$ represent the marginals for the posterior probability $w_{ig\ell}$ and for the joint density $f_{ig\ell}$, respectively.

As it is typical in finite mixture models, equation (\ref{eq:score3}) can be solved analytically to give the updates  
\[
\hat \pi_{g\ell}^{(r)} = \frac{\sum_{i=1}^n w_{ig\ell}^{(r)}}{n},
\]
while the remaining model parameters may be updated by using standard Newton-type algorithms.  The E- and the M-step of the algorithm are alternated until convergence, that is, until the (relative) difference between two subsequent likelihood values is smaller than a given quantity $\varepsilon > 0$. Given that this criterion may indicate lack of progress rather than true convergence, see eg \cite{karl2001}, and the log-likelihood may suffer from multiple local maxima, we usually suggest to start the algorithm from several different starting points. In all the following analyses, we used B=50 starting points.
Also, as it is typically done when dealing with finite mixtures, the number of locations $K_1$ and $K_2$ is treated as fixed and known. The algorithm is run for varying $(K_1, K_2)$ combinations and the optimal solution is chosen via standard model selection techniques, such as AIC, \citep{Akaike1973} or BIC \citep{Schwarz1978}.

Standard errors for model parameter estimates are obtained at convergence of the EM algorithm by the standard sandwich formula \citep{White1980, Royall1986}. This leads to the following estimate of the covariance matrix of model parameters:
\begin{align*}
\label{eq:swd}
\widehat{\rm Cov(\hat{\bUpsilon})} = \tbf I_o(\hat{\bUpsilon})^{-1} \widehat{\mbox{Cov}(\tbf S)} 
\tbf I_o(\hat{\bUpsilon})^{-1}, 
\end{align*}
where $\tbf I_o(\hat{\bUpsilon})$ denotes the observed information matrix which is computed via the Oakes' formula \citep{oak1999}. Furthermore, $\tbf S$ denotes the score vector evaluated at $\hat \bUpsilon$ and $\widehat{\mbox{Cov}(\tbf S)} = \sum_{i=1}^n  \tbf S_i(\hat{\bUpsilon}) \tbf S_i^\prime(\hat{\bUpsilon})$ denotes the estimate for the covariance matrix of the score function $\rm{Cov}(\tbf S)$, with $\tbf S_i$ being the individual contribution to the score vector.

\section{Sensitivity analysis: definition of the index} \label{sec:6}
The proposed bi-dimensional finite mixture model allows to account for possible effects of non-ignorable dropouts on the primary outcome of interest. However, as highlighted by \cite{mol2008}, for every MNAR model there is a corresponding MAR counterpart that produces exactly the same fit to observed data. This is due to the fact that the MNAR model is fitted by using the observed data only and it implicitly assumes that the distribution of the missing responses is identical to that of the observed ones. Further, the dependence between the longitudinal response (observed and missing) and the dropout indicator which is modeled via the proposed model specification is just one out of several possible choices.
Therefore, rather than relying on a single (possibly misspecified) MNAR model and in order evaluate how maximum likelihood estimates for the longitudinal model parameters  are influenced by the hypotheses about the dropout mechanism, a sensitivity analysis is always recommended. 

In this perspective, most of the available proposals focus on Selection or Pattern Mixture Model specifications \citep{Little1995}, while few proposals are available for shared random coefficient models. A notable exception is the proposal by \cite{cre2010}. Here, the authors considered a sensitivity parameter in the model and studied how model parameter estimates vary when the sensitivity parameter is forced to move away from zero. 
Looking at \emph{local} sensitivity, \cite{trox2004} developed an index of local sensitivity to non-ignorability (ISNI) via a first order Taylor expansion, with the aim at describing the "geometry" of the likelihood surface in a neighborhood of a MAR solution. 
Such an index was further extended by \cite{ma2005}, \cite{Xie2004}, and \cite{Xie2008} to deal with the general case of $q$-dimensional ($q >1$) non ignorability  parameters by considering an $L_{2}$ to summarize the impact of a unit change in its elements. 
An $L_{1}$ norm was instead considered by \cite{Xie2012}, while \cite{Gao2016} further extended the ISNI definition by considering a higher order Taylor expansion. 
In the context of joint  models for longitudinal responses and (continuous) time to event data, \cite{viv2014} proposed a relative index based on the ratio between the ISNI and a measure of its variability under the MAR assumption. 

Due to the peculiarity of the proposed model specification, to specify a index of sensitivity to non-ignorability we proceed as follows.
As before, let $\blambda = (\lambda_{11}, \dots, \lambda_{K_1K_2})$ denote the vector of non ignorability parameters and let $\blambda={\bf 0}$ correspond to a MAR model. Also, let $\bxi = (\xi_{11}, \dots, \xi_{K_1K_2})$ denote the vector of all logit transforms defined in equation \eqref{pi:equation} and let $\bxi_0$ correspond to a MAR model. That is, $\bxi_0$ has elements
\[
\xi_{gl} = \alpha_{g\star} + \alpha_{\star \ell}, \quad g=1, \dots, K_1, \ell=1, \dots, K_2.
\]
Both vectors $\blambda$ and $\bxi$ may be interchangeably considered as non-ignorability parameters in the proposed model specification, but to be coherent with the definition of the index, we will use $\blambda$ in the following. 
Last, let us denote by $\hat{\boldsymbol \Phi}(\blambda)$ the maximum likelihood estimate for  model parameters in the longitudinal data model, conditional on a given value for the sensitivity parameters $\blambda$. 
 
The \emph{index of sensitivity to non-ignorability} may be derived as
\begin{equation}
\label{eq:ISNIbase}
ISNI_{\bPhi}=\left.\frac{\partial \hat{\bPhi}(\blambda)} {\partial \blambda} \right|_{\bl \Phi(\bl 0)} \simeq -  \left(\left.\frac{\partial^{2} \ell(\bPhi, \bPsi, \bpi)}{\partial \bPhi \bPhi^{\prime}}\right|_{\bl \Phi(\bl 0)}\right)^{-1}  \left. \frac{\partial^{2} \ell(\bPhi, \bPsi, \bpi)}{\partial \bPhi \blambda}\right|_{\bl \Phi(\bl 0)}.
\end{equation}

Based on the equation above, it is clear that the \textit{ISNI} measures the displacement of model parameter estimates from their MAR counterpart, in the direction of $\blambda$, when we move far from $\blambda = \b 0$.
Following similar arguments as those detailed by \citep{Xie2008}, it can be shown that the following expression holds: 
\begin{align*}
\label{eq:ISNIapprox}
\hat \bPhi(\blambda)=\hat \bPhi({\bf 0})+ISNI_{\bPhi}\blambda;
\end{align*}
that is, the ISNI may be also interpreted as the linear impact that changes in the elements of $\blambda$ have on $\hat \bPhi$.

It is worth to highlight that, $ISNI_{\bPhi}$ denotes a matrix with $D$ rows and $(K_{1}-1)(K_{2}-1)$ columns representing the effect each element in $\blambda$ has on the $D$ elements in $\bPhi$. That is, the proposed formulation of the index leads to a matrix rather than a scalar or a vector as in the original formulations. 
In this respect, to derive a global measure of local sensitivity for the parameter estimate $\hat \Phi_d$ when moving far from the MAR assumption, for $ d =1, \dots, D,$ a proper summary of the corresponding row in the \textit{ISNI} matrix, say $ISNI_{\Phi_d}$, needs to be computed.

\section{Analysis of the Leiden 85+ data}
\label{sec:7}
In this section, the bi-dimensional finite mixture model is applied to the analysis of the Leiden 85+ study data. We aim at understanding the effects for a number of covariates on the dynamics of cognitive functioning in the elderly, while controlling for potential bias in the parameter estimates due to possible non-ignorable dropouts.
First, we provide a description of the available covariates in section \ref{sec_preliminary} and describe the sample in terms of demographic and genetic characteristics of individuals participating in the study. Afterwards, we analyze the joint effect of these factors on the dynamics of the (transformed) MMSE score. Results are reported in sections \ref{sec_MARmodel}--\ref{sec_MNARmodel}-. Last in section \ref{sec_isni}, a sensitivity analysis is performed to give insight on changes in parameters estimates when we move far from the MAR assumption. Two scenarios are investigated and results reported.

\subsection{Preliminary analysis}\label{sec_preliminary}
We start the analysis by summarizing in Table \ref{tabII} individual features of the sample of subjects participating in the Leiden 85+ study, both in terms of covariates and MMSE scores, conditional on the observed pattern of participation. That is, we distinguish between those individuals who completed the study and those who did not. As highlighted before, subjects who present incomplete information are likely to leave the study because of poor health conditions and this questions weather the analysis based on the observed data only may lead to biased results. 

By looking at the overall results, we may observe that the $64.88\%$ of the sample has a low level of education and females represent the $66.73\%$ of whole sample. As regards the \textit{APOE} genotype, the most referenced category is obviously $APOE_{33}$ $(58.96 \%)$, far from $APOE_{34-44}$ $(21.08\%)$ and $APOE_{22-23}$ $(17.74 \%)$, while only a very small portion of the sample ($2.22\%$) is characterized by $APOE_{24}$.
Last, we may notice that more than half of the study participants ($50.83\%$) leave the study before the scheduled end. This proportion is  relatively higher for participants with low level of education ($52.71\%$), for males ($58.89\%$), and for those in the $APOE_{34-44}$ group ($61.40\%$).

\begin{table}[htb]
\caption{Leiden 85+ Study: demographic and genetic characteristics of participants}
\label{tabII}
\begin{center}
\begin{tabular}{l c c c }
\hline
Variable                 & Total  & Completed (\%)  &  Did not complete (\%)     \\ \hline
\textbf{Gender }         &        		&        		    &                              \\
Male                     & 180 (33.27)	& 74 (41.11) 	    &    106 (58.89)          \\
Female                   & 361 (66.73)  & 192 (53.19)     &    169 (46.81)           \\
\textbf{Education}     &        &            	    &                     \\
Primary                  & 351 (64.88)  & 166 (47.29)     &   185 (52.71)           \\
Secondary                & 190 (35.12)  & 100 (52.63)     &   90  (47.37)        \\
\textbf{APO-E}           &        &                 &                            \\
22-23                    &  96 (17.74)  &    54 (56.25)   &   42  (43.75)          \\
24                       &  12  (2.22)  &    6  (50)      &    6 (50)           \\
33                       & 319 (58.96)  &   162 (50.78)   &  157 (49.22)        \\
34-44                    & 114 (21.08)  &    44 (38.60)   &   70  (61.40)         \\
 \hline
Total                    & 541 (100)  &    266 (49.17)   &   275  (50.83)         \\
\end{tabular}
\end{center}
\end{table}

\begin{center}
\begin{figure}[h!]
\caption{Leiden 85+ Study: mean of MMSE score stratified by age, and gender(a), educational level (b), APOE (c)}
\centerline{\includegraphics[scale=0.7]{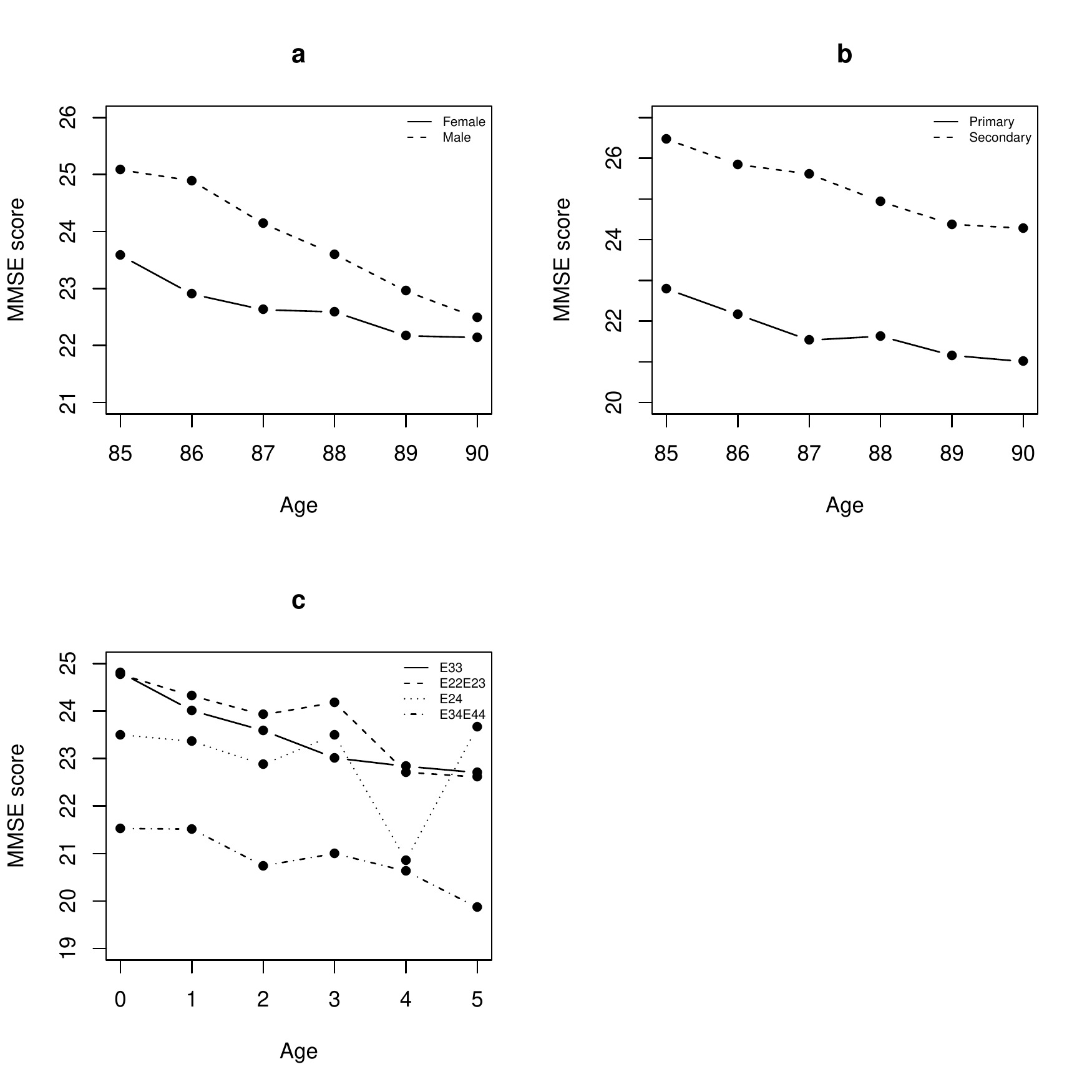}}
\label{fig:plot_cov}
\end{figure}
\end{center}

Figure \ref{fig:plot_cov} reports the evolution of the mean MMSE over time stratified by the available covariates. As it is clear, cognitive impairment is higher for males than females, even if the differences seem to decrease with age, maybe due to a direct age effect or a to differential dropout by gender (Figure \ref{fig:plot_cov}a). By looking at Figure \ref{fig:plot_cov}b, we may also observe that participants with higher education are less cognitively impaired at the begining of the study, and this difference remains persistent over the analyzed time window. Rather than only a direct effect of education, this may suggest differential socio-economic statuses being associated to differential levels of education. Last, lower MMSE scores are observed for $APOE_{34-44}$, that is when allele $\epsilon4$, which is deemed to be a risk factor for dementia, is present. The irregular pattern for $APOE_{24}$ may be due to the sample size of this group (Figure \ref{fig:plot_cov}c). 

\subsection{The MAR model}\label{sec_MARmodel}
We start by estimating a MAR model, based on the assumption of independence between the longitudinal and the dropout process. In terms of equation (\ref{eq:log-likelihooddouble}), this is obtained by assuming $\pi_{g\ell}=\pi_{g\star}\pi{\star \ell}$, for $g=1,\dots,K_{1}$ and $\ell=1,\dots,K_{2}$. Alternatively, we can derive it by fixing $\blambda={\bf 0}$ in equation (\ref{pi:equation}) or $M=1$ in eq. (\ref{eq:pidecomp}). To get insight on the effects of demographic and genetic features on the individual dynamics of the MMSE score, we focused on the following model specification:
\begin{align*}
\left\{
\begin{array}{ccc}
Y_{it} \mid \textbf x_{it},b_{i1} \sim {\rm N}(\mu_{it}, \sigma^{2}) \\
R_{it} \mid \textbf v_{it},b_{i2} \sim {\rm Bin}(1, \phi_{it})
\end{array}
\right.
\end{align*}
The canonical parameters are defined by the following regression models:
\begin{eqnarray*}
\mu_{it}&=&(\beta_{0}+b_{i1})+\beta_{1}\, (Age_{it}-85)+\beta_{2}\, Gender_{i}+\beta_{3}\, Educ_{it}+ \\&+&\beta_{4}\, APOE_{22-23}+\beta_{5}\, APOE_{24}+\beta_{6}\, APOE_{34-44}, \\[2mm]
{\rm logit}(\phi_{it})&=&(\gamma_{0}+b_{i2})+\gamma_{1}\, (Age_{it}-85)+\gamma_{2}\, Gender_{i}+\gamma_{3}\, Educ_{it}+\\
&+&\gamma_{4}\, APOE_{22-23}+\gamma_{5}\, APOE_{24}+\gamma_{6}\, APOE_{34-44}.
\end{eqnarray*}
As regards the response variable, the transform $Y_{it} = \log[1+ (30 - \mbox{MMSE}_{it})]$ was adopted as it is nearly optimal in a Box-Cox sense.

Both a parametric and a semi-parametric specification of the random coefficient distribution were considered. In the former case, Gaussian distributed random effects were inserted into the linear predictors for the longitudinal response and the dropout indicator. In the latter case, for each margin, the algorithm was run for varying number of locations and the solution corresponding to the lowest BIC index was retained, leading to the selection of $K_1 = 5$ and $K_2 = 3$ components for the longitudinal and the dropout process, respectively. Estimated parameters, together with corresponding standard errors are reported in Table \ref{tabmar}.

\begin{table}

\caption{Leiden 85+ Study: MAR models. Maximun likelihood estimates, standard errors, log-likelihood, and BIC value}
\label{tabmar}
\begin{center}
\begin{tabular}{l|l r r r r  }
\hline
Process &                 & \multicolumn{2}{c}{Semi-parametric}& \multicolumn{2}{c}{Parametric} \\
    &   Variable        & Coeff.     & Std. Err. &  Coeff.        & Std. Err.   \\ \hline\hline
	&	\textit{Intercept}	    &	1.686	 &		    &	1.792	&	0.050 \\
	&	\textit{Age}	            &	0.090	 &	0.008	&	0.089	&	0.005 \\
	&	\textit{Gender}	        &	-0.137	 &	0.042	&	-0.085	&	0.066 \\
	&	\textit{Educ}	        &	-0.317	 &	0.068	&	-0.623	&	0.065 \\
Y	&	$APOE_{22-23}$  &	0.062	 &	0.072	&	0.056	&	0.083 \\
	&	$APOE_{24}$	    &	-0.105	 &	0.062	&	0.096	&	0.211 \\
	&	$APOE_{34-44}$  &	0.347	 &	0.060	&	0.369	&	0.079 \\
	&	$\sigma_{y}$    &	0.402	 &		    &	0.398	&	\\
	&	$\sigma_{b_{1}}$&	0.696	 &		    &	0.684	&	
\\ \hline
	&	\textit{Intercept}	    &	-11.475	 &		    &	-3.877	&	0.520 \\
	&	\textit{Age}				&	2.758	 &	0.417	&	0.526	&	0.131 \\
	&	\textit{Gender}			&	0.559	 &	0.467	&	0.656	&	0.218 \\
	&	\textit{Educ}			&	-2.162	 &	0.772	&	-0.486	&	0.212 \\
R	&	$APOE_{22-23}$	&	0.476	 &	0.409	&	-0.246	&	0.252 \\
	&	$APOE_{24}$		&	-0.026	 &	0.939	&	0.131	&	0.618 \\
	&	$APOE_{34-44}$	&	0.805	 &	0.461	&	0.565	&	0.237 \\
	&	$\sigma_{b_{2}}$&	5.393	 &		    &	1.525	&	\\ \hline
&	$\log L$        &   -2685.32 &		    & -2732.84	&	\\
	&	BIC	            &	5534.26 &		    & 5572.67	&	\\ \hline
\end{tabular}
\end{center}
\end{table}

By looking at the results, few findings are worth to be discussed. First, the estimates obtained via either the parametric or the semi-parametric approach are quite similar when we consider the longitudinal process. That is, $\log\left[1+(30-MMSE)\right]$ increases (and MMSE decreases) with age. A significant gender effect can be observed, with males being less impaired (on average) than females. Furthermore, a strong protective effect seems to be linked to socio-economic status in early life as it may be deduced from the significant and negative effect of higher educational levels. Table \ref{tabmar} also highlights how $APOE_{34-44}$ represents a strong risk factor, with a positive estimate on the adopted response scale and, therefore, a negative effect on the MMSE. Only few differences may be highlighted when comparing the estimates obtained under the parametric and the semi-parametric approach for the longitudinal data process. In particular, these differences are related to the gender effect, which is not significant in the parametric model, and the effect of higher education, which is much higher under the parametric specification. This differences may be possibly due to the discrete nature of the random effect distribution in the semi-parametric case, which may lead to partial aliasing with the time-constant covariates. 

When the dropout process is considered, we may observe that the results are \emph{qualitatively} the same, but the size of parameter estimates is quite different. This could be due, at least partially, to the different scale of the estimated random coefficient distribution, with $\sigma_{b_{2}}=5.393$ and $\sigma_{b_2} = 1.525$ in the semi-parametric and in the parametric model, respectively. As it is clear, in the semi-parametric case, the estimated intercepts are quite higher than those that can be predicted by a Gaussian distribution and this leads to inflated effects for the set of observed covariates as well. 
However, if we look at the estimated dropout probabilities resulting either from the semi-parametric or the parametric models, these are very close to each other, but for few extreme cases which are better recovered by the semi-parametric model.

\subsection{The MNAR model}\label{sec_MNARmodel}
To provide further insight into the effect of demographic and genetic factors on the MMSE dynamics, while considering the potential non-ignorability of the dropout process, we fitted both a uni-dimensional and a bi-dimensional finite mixture model. For the former approach, we run the estimation algorithm for $K = 1, \dots, 10$ and retained the optimal solution according to the BIC index.
This corresponds to a model with $K = 5$ components.
Similarly, for the proposed bi-dimensional finite mixture model, we run the algorithm for $K_1 = 1, \dots, 10$ and $K_2 = 1, \dots, 5$ components and, as before, we retained as optimal solution that with the lowest BIC. That is, the solution with $K_1 = 5$ and $K_2=3$ components for the longitudinal and the dropout process, respectively. 
This result is clearly coherent with that obtained by marginal modeling the longitudinal response and the dropout indicator. 
Parameter estimates and the corresponding standard errors for both model specifications are reported in Table \ref{tabmnar}.

\begin{table}
\caption{Leiden 85+ Study: MNAR models. Maximun likelihood estimates, standard errors, log-likelihood, and BIC value}\label{tabmnar}
\begin{center}
\begin{tabular}{l|l  r r r r }
\hline
Process & & \multicolumn{2}{c}{Semipar. ``Uni-dim."} & \multicolumn{2}{c}{Semipar.``Bi-dim."} \\
        & Variable 		&  Coeff. & Std. Err.  &  Coeff. & Std. Err.  \\ \hline\hline
	&	Intercept		&	1.682	&		&	1.687	&	 \\
	&	Age				&	0.094	&	0.007	&	0.094	&	0.007 \\
	&	Gender			&	-0.129	&	0.048	&	-0.135	&	0.039 \\
Y	&	Educ			&	-0.31	&	0.051	&	-0.317	&	0.050 \\
	&	APOE$_{22-23}$	&	0.091	&	0.061	&	0.086	&	0.058 \\
	&	APOE$_{24}$		&	-0.098	&	0.055	&	-0.099	&	0.056 \\
	&	APOE$_{34-44}$	&	0.345	&	0.050	&	0.344	&	0.051 \\
	&	$\sigma_{y}$	&	0.402	&			&	0.402	&	 \\
	&	$\sigma_{b_{1}}$&	0.701	&			&	0.699	&	\\ 
\hline 
	&	Intercept		&	-3.361	&			&	-10.767	&	\\
	&	Age				&	0.367	&	0.037	&	2.406	&	0.384 \\
	&	Gender			&	0.504	&	0.147	&	1.061	&	0.850 \\
R	&	Educ			&	-0.200	&	0.151	&	-1.646	&	0.530 \\
	&	APOE$_{22-23}$	&	-0.090	&	0.199	&	0.481	&	1.090 \\
	&	APOE$_{24}$		&	-0.148	&	0.508	&	-0.334	&	0.647 \\
	&	APOE$_{34-44}$	&	0.541	&	0.174	&	1.365	&	0.745 \\
	&	$\sigma_{b_{2}}$&	0.577	&			&	4.891	&	\\
	&	$\sigma_{b_{1},b_{2}}$	&	0.349 		&			&	0.985	&	\\
	&	$\rho_{b_{1},b_{2}}$	&	0.863		&			& 	0.288		&	\\ 
\hline \hline
&	$\log L$					&	-2686.902	&			&	-2660.391	&	\\
	&	BIC						&	5537.433	&			&	5534.758	&	\\ \hline
\end{tabular}
\end{center}
\end{table}

When looking at the estimated parameters for the longitudinal data process and at their significance (left panel in the table), we may conclude that estimates are coherent with those obtained in the MAR analysis. A small departure can be observed for the effect of age and gender. Males and patients with high education tend to be less cognitively impaired when compared to the rest of the sample, while subjects carrying $\epsilon4$ alleles, that is with category $APOE_{34-44}$, present a steeper increase in the observed response, e.g. a steeper decline in MMSE values. Focusing the attention on the dropout process, we may observe that age, gender and $APOE_{33-34}$ are all positively related with an increased dropout probability. That is, older men carrying $\epsilon4$ alleles are more likely to leave the study prematurely than younger females carrying $\epsilon3$ alleles.

By comparing the estimates obtained under the uni- and the bi-dimensional finite mixture model, it seems that the above results hold besides the chosen model specification. The only remarkable difference is in the estimated magnitude of the effects for the dropout process and for the random coefficient distribution. For the bi-dimensional finite mixture model, we may observe a stronger impact of the covariates on the dropout probability. However, as for the univariate model described in section \ref{sec_MARmodel}, this result is likely due to the estimated scale with an intercept value which is much lower under the bi-dimensional specification than under the uni-dimensional one. 
Further, under the uni-dimensional model specification, the Gaussian process for the longitudinal response may have a much higher impact on the likelihood function when compared to the Bernoulli process for the dropout indicator. As a result, the estimates for component-specific locations and the corresponding variability in the dropout equation substantially differ when comparing the uni-dimensional and the univariate model. In the uni-dimensional model, the estimated correlation is quite high due to reduced variability of the random coefficients in the dropout equation, while this is substantially reduced in the bi-dimensional case.

We also report in Table \ref{tabprob} the estimated random intercepts for the longitudinal and the dropout process, together with the corresponding conditional distribution i.e. $\pi_{\ell \mid g} = \Pr(b_{i2}=\zeta_{2l} \mid b_{i1}=\zeta_{1g}=$. When focusing on the estimated locations in the longitudinal data process, that is $\zeta_{1g}$, we may observe higher cognitively impairment when moving from the first to the latter mixture component. On the other hand, for the dropout process, estimated locations, $\zeta_{2\ell}$, suggest that higher components correspond to a higher chance to dropout from the study. 
When looking at the estimated conditional probabilities, we may observe a link between lower (higher) values of $\zeta_{1g}$ and lower (higher) values of $\zeta_{2\ell}$. That is, participants with better cognitive functioning (i.e. with lower response values) are usually characterized by a lower probability of dropping out from the study. On the contrary, cognitively impaired participants (i.e. with higher response values) present a higher chance to dropout prematurely from the study, even if there is still some overlapping between the second and the third component in the dropout profile.

\begin{table}[h!]
\caption{Maximun likelihood estimates and conditional distribution for the random parameters}\label{tabprob}
\centering
\begin{tabular}{l | c c c | c}
&     \multicolumn{3}{c}{$\zeta_{2\ell}$} & \\ \hline
$\zeta_{1g}$	&	-15.053	&	-8.701	&	-3.378	&	\\ \hline\hline
0.519	&	0.865	&	0.090	&	0.045	&	1 \\
1.065	&	0.585	&	0.170	&	0.245	&	1 \\
1.681	&	0.573	&	0.227	&	0.199	&	1 \\
2.297	&	0.467	&	0.289	&	0.244	&	1 \\
2.905	&	0.144	&	0.364	&	0.492	&	1 \\  \hline
Tot.	&	0.528	&	0.229	&	0.243	&	1
\end{tabular}
\end{table}

Looking at the parameter estimates obtained through the MNAR model approach, we may observe a certain degree of correlation between the random effects in the two equations. This suggests the presence of a potential non-ignorable dropout process affecting the longitudinal outcome. However, such an influence cannot be formally tested, as we may fit the proposed model to the observed data only and derive estimates on the basis of strong assumptions on the behavior of missing responses. Therefore, it could be of interest to verify how assumptions on the missing data mechanism can influence parameter estimates.

\subsection{Sensitivity analysis: results}\label{sec_isni}
To investigate the robustness of inference with respect to the assumptions on the missing data mechanism, we computed the matrix $ISNI_{\bPhi}$ according to formulas provided in equation  \eqref{eq:ISNIbase}. 
For each model parameter estimates $\hat \Phi_d$, we derived a global measure of its sensitivity to the MAR assumption by computing the norm, the minimum and the maximum of $\lvert ISNI_{\hat \Phi_d}\rvert$ and its ratio to the corresponding standard error estimates from the MAR model.

\begin{table}[h!]
\caption{MAR model estimates: ISNI norm, minimum and maximum (in absolute values), and ratio to the corresponding standard error.}\label{tabISNI}
\centering
\scalebox{0.85}{
\begin{tabular}{l | c c c c c c c}
\hline
	&	se	&	ISNI	&	norm(ISNI)/se	&	$\lvert ISNI\rvert$	&	min$\lvert ISNI\rvert$/se	&	ISNI	&	max$\lvert ISNI\rvert$/se \\
	
Variable	&		&	(norm)	&		&	(min)	&		&	(max)	&	\\ \hline
$\bzeta_{11}$	&	0.117	&	0.0414	&	0.354	&	0.0014	&	0.012	&	0.0204	&	0.174	\\
$\bzeta_{12}$	&	0.074	&	0.0580	&	0.784	&	0.0016	&	0.022	&	0.0303	&	0.409	\\
$\bzeta_{13}$	&	0.074	&	0.044	&	0.595	&	0.0002	&	0.003	&	0.0255	&	0.345	\\
$\bzeta_{14}$	&	0.083	&	0.1044	&	1.258	&	0.0005	&	0.006	&	0.0527	&	0.635	\\
$\bzeta_{15}$	&	0.071	&	0.0088	&	0.124	&	0.0009	&	0.013	&	0.0045	&	0.063	\\
\textit{Age}	&	0.008	&	0.0089	&	1.113	&	0.0001	&	0.013	&	0.0054	&	0.675	\\
\textit{Gender}	&	0.042	&	0.0058	&	0.138	&	0.0003	&	0.007	&	0.0028	&	0.067	\\
\textit{Educ}	&	0.068	&	0.0075	&	0.110	&	0.0001	&	0.001	&	0.004	&	0.059	\\
$APOE_{22-23}$	&	0.072	&	0.0111	&	0.154	&	0.0001	&	0.001	&	0.0074	&	0.103	\\
$APOE_{24}$	&	0.062	&	0.0123	&	0.198	&	0.0005	&	0.008	&	0.0051	&	0.082	\\
$APOE_{34-44}$	&	0.06	&	0.012	&	0.200	&	0.0009	&	0.015	&	0.0061	&	0.102	\\
$\sigma_{y}$	&	0.194	&	0.1123	&	0.579	&	0.0001	&	0.001	&	0.0824	&	0.425	\\

 \hline
\end{tabular}
}
\end{table}

{By looking at the results reported in Table \ref{tabISNI}, we may observed that, as far as fixed model parameters are concerned, the global indexes we computed to investigate how they vary when moving far from the MAR assumption are all quite close to zero. The only remarkable exception is for the \textit{age} variable. In this case, the \textit{ISNI} seems to take slightly higher values and this is particularly evident when focusing on the standardized statistics. Higher \textit{ISNI} values may be also observed when looking at the random intercepts. However, this represents an expected results being this parameters  connected (even if indirectly) to the missingness process.}

\begin{figure}
\caption{Leiden 85+ Study: Sensitivity analysis according to Scenario 1}
\centerline{\includegraphics[scale=0.55]{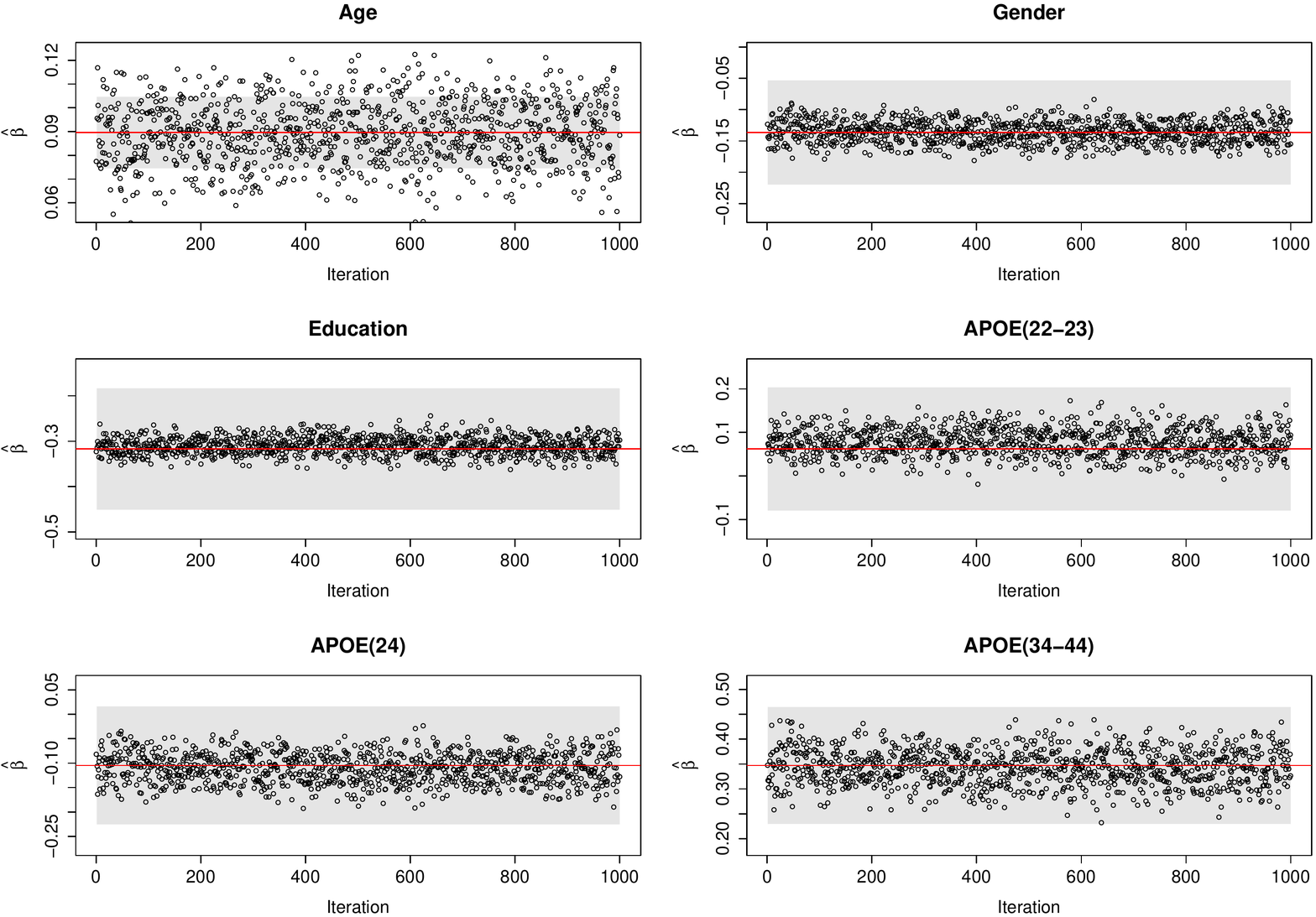}}
\label{FigureScenario1}
\end{figure}

\begin{figure}
\caption{Leiden 85+ Study: Sensitivity analysis according to Scenario 1}
\centerline{\includegraphics[scale=0.55]{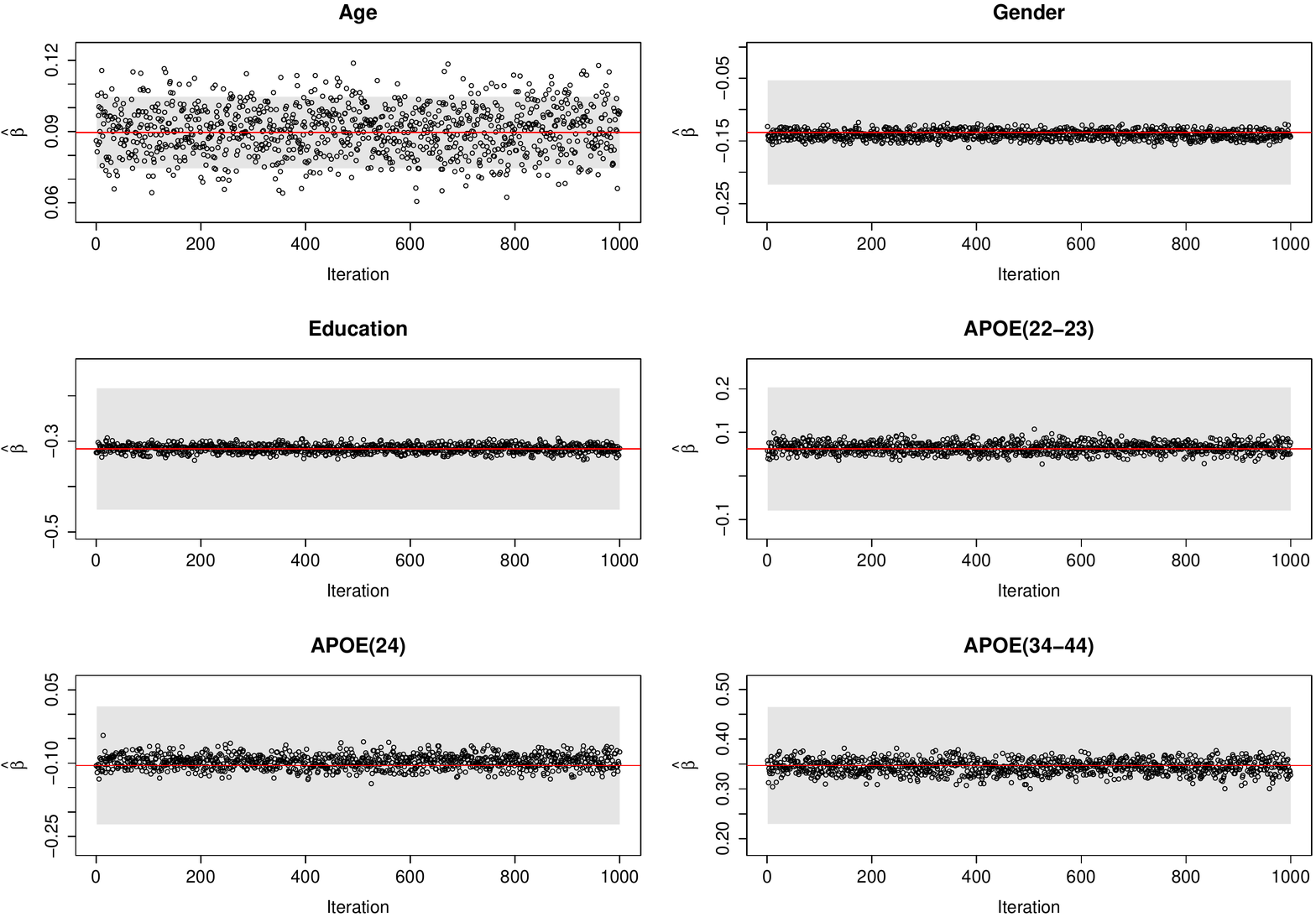}}
\label{FigureScenario2}
\end{figure}

To further study the potential impact that assumptions on the missing data generating mechanism may have on the parameters of interest, we may analyze how changes in $\blambda$ parameters affect the vector $\hat \bPhi$. In this respect, we considered the following two scenarios.
\begin{itemize}
\item[Scenario 1]
We simulated $B=1000$ values for each element in $\blambda$ from a uniform distribution, $\lambda_{g \ell}(b) \sim {\rm U}(-3,3)$ for $g=1,\dots,K_{1}-1$ and $\ell=1,\dots,K_{2}-1$. Then, based on the simulated values, we computed 
\[
\hat \bPhi(b)=\hat \bPhi({\bf 0})+ISNI_{\bPhi} \times \blambda(b).
\]
\item[Scenario 2]
We simulated $B=1000$ values for a scale constant $c$ from a uniform distribution, $c(b) \sim {\rm U}(-3,3)$. Then, based on the simulated values, we computed 
\[
\xi_{g \ell}(b) =\xi({\bf 0}) + c(b) \hat{\lambda}_{g \ell}, \qquad g=1,\dots,K_{1}-1, \ell=1,\dots,K_{2}-1,
\]
where $\hat{\lambda}_{g\ell}$ denotes the maximum likelihood estimate of $\lambda_{g\ell}$ under the MNAR model. This scenario allows us to consider perturbations in the component specific masses, while preserving the overall dependence structure we estimated through the proposed MNAR model. That is, this allows us to link changes in the longitudinal model parameters with increasing (respectively decreasing) correlation between the random coefficients in the two profiles of interest. The  corresponding (approximated) parameter estimates are computed as
\[
\hat \bPhi(b)=\hat \bPhi({\bf 0})+ISNI_{\bPhi}\blambda(b),
\]
where $\blambda(b)=c(b) \hat{\blambda}$.
\end{itemize}
{The first scenario is designed to study the general sensitivity of parameter estimates. That is, we aim at analyzing how model parameter estimates vary when random changes in $\blambda$ (in any direction) are observed. The second scenario starts from the estimated  pattern of dependence between the random intercepts in the longitudinal and the missing data models and try to get insight on the changes in parameter estimates that could be registered in the case the correlation increases (in absolute value) with respect to the estimated one. 
In Figures \ref{FigureScenario1} and \ref{FigureScenario2} we report 
parameter estimates derived under Scenario 1 and Scenario 2, respectively. The red line and the grey bands in each graph correspond to the point and the $95\%$ interval estimates of model parameters under the MAR assumption.}

When focusing Figure \ref{FigureScenario1} (Scenario 1), it can be easily observed that only the parameter associated to the \textit{age} variable is slightly sensitive to changes in the assumptions about the ignorability of the dropout process.
All the other estimates remain quite constant and, overall, within the corresponding $95\%$ MAR confidence interval. 
No particular patterns of dependence/correlation between the random coefficients can be linked to points outside the interval for the estimated effect of $age$.  
{Rather, we observed that strong \emph{local} changes in the random coefficient probability matrix may cause positive (respectively  negative) changes in the \textit{age} effect. In particular, changes in the upper left or intermediate right components, that is components with low values of both random coefficients (first case) or with high values for $\zeta_{2\ell}$ and intermediate values for $\zeta_{ig}$, respectively. 
}
Overall, the relative frequency of points within the corresponding MAR confidence interval is equal to $0.737$, which suggests a certain sensitivity to assumptions regarding ignorability of the dropout process, even though estimates always remain within a reasonable set.

When focusing on Figure \ref{FigureScenario2} (Scenario 2), we may observe that changes in the parameter estimates are more clearly linked to correlation between the random effects in the two profiles. As for the former scenario, a slight  sensitivity to departures from the MAR assumption is observed for the \textit{age} variable only. 
In this case, the relative frequency of points within the corresponding MAR confidence interval for the \textit{age} effect is equal to $0.851$, which suggests a lower sensitivity to assumptions on the ignorability of the dropout process, when compared to the one observed under Scenario 1. In this case, high positive correlation between the random coefficients leads to MAR estimates that are lower than the corresponding MNAR counterparts. On the other hand, high negative correlation leads to MAR estimates that tend to be higher than the MNAR counterparts.

The proposed approach for sensitivity analysis could be seen as a particular version of local influence diagnostics developed in the context of regression models to check for influential observations by perturbations in individual-specific weights; see eg \cite{Jans2003} and \cite{rakh2016,rakh2017} for more recent developments. Here, rather than perturbing individual observations, we perturb weights associated to the group of subjects allocated to a given component. Obviously, a \emph{global} influence approach could be adopted as well, for example by looking at the \emph{mean score} approach detailed in \cite{whit2017}.

\section{Conclusions}
\label{sec:8}
We defined a random coefficient based dropout model where the association between the longitudinal and the dropout process is modeled through discrete, outcome-specific, latent effects. A bi-dimensional representation for the random coefficient distribution was used and a (possibly) different number of locations in each margin is allowed. A full probability matrix connecting the locations in a margin to those in the other one was considered. The main  advantage of this flexible representation for the random coefficient distribution is that the resulting MNAR model properly nests a model where the dropout mechanism is non ignorable. This allows us to consider a (local) sensitivity analysis, based on the ISNI index, to check changes in model parameter estimates as we move far from the MAR assumption.
The data application showed good robustness of all model parameter estimates. A slight sensitivity to assumptions on the missing data generating mechanism was only observe for the $age$ effect which, however, is always restricted to a reasonable set.

\section*{Acknowledgement}
We gratefully acknowledge Dr Ton de Craen and Dr Rudi Westendorp of the Leiden University Medical Centre, for kindly providing the analyzed data.

%
%

\begin{thebibliography}{44}
\providecommand{\natexlab}[1]{#1}
\providecommand{\url}[1]{\texttt{#1}}
\expandafter\ifx\csname urlstyle\endcsname\relax
  \providecommand{\doi}[1]{doi: #1}\else
  \providecommand{\doi}{doi: \begingroup \urlstyle{rm}\Url}\fi

\bibitem[Aitkin and Alf\'o(2006)]{ait2003}
M.~Aitkin and M.~Alf\'o.
\newblock Variance component models for longitudinal count data with baseline
  information: epilepsy data revisited.
\newblock \emph{Statistics and Computing}, 16:\penalty0 231--238, 2006.

\bibitem[Akaike(1973)]{Akaike1973}
H.~Akaike.
\newblock {Information theory and an extension of the maximum likelihood
  principle}.
\newblock In B.~N. Petrov and F.~Csaki, editors, \emph{Second International
  Symposium on Information Theory}, pages 267--281. Akad\'{e}miai Kiado, 1973.

\bibitem[Alf\'o and Maruotti(2009)]{alf2009}
M.~Alf\'o and A.~Maruotti.
\newblock A selection model for longitudinal binary responses subject to
  non-ignorable attrition.
\newblock \emph{Statistics in Medicine}, 28:\penalty0 2435--2450, 2009.

\bibitem[Alf\'o and Rocchetti(2013)]{alf2012}
M.~Alf\'o and I.~Rocchetti.
\newblock A flexible approach to finite mixture regression models for
  multivariate mixed responses.
\newblock \emph{Statistics and Probability Letters}, 83:\penalty0 1754--1758,
  2013.

\bibitem[Bartolucci and Farcomeni(2015)]{bart2015}
F.~Bartolucci and A.~Farcomeni.
\newblock A discrete time event-history approach to informative drop-out in
  mixed latent markov models with covariates.
\newblock \emph{Biometrics}, 71:\penalty0 80--89, 2015.

\bibitem[Beunckens et~al.(2008)Beunckens, Molenberghs, Verbeke, and
  Mallinckrodt]{Beunc2008}
C.~Beunckens, G.~Molenberghs, G.~Verbeke, and C.~Mallinckrodt.
\newblock A latent-class mixture model for incomplete longitudinal gaussian
  data.
\newblock \emph{Biometrics}, 64:\penalty0 96--105, 2008.

\bibitem[Carlin and Luois(2000)]{car2000}
B.P. Carlin and T.A. Luois.
\newblock \emph{Bayes and Empirical Bayes Methods for Data Analysis}.
\newblock Chapman \& all, 2000.

\bibitem[Chamberlain(1984)]{cham1984}
G.~Chamberlain.
\newblock Panel data.
\newblock In Z.~Griliches and M.D. Intriligator, editors, \emph{Handbook of
  Econometrics, Volume II}, chapter~22, pages 1247--1318. Elsevier Science
  Publishers BV, Amsterdam, 1984.

\bibitem[Creemers et~al.(2010)Creemers, Hens, Aerts, Molenberghs, Verbeke, and
  Kenward]{cre2010}
A.~Creemers, N.~Hens, M.~Aerts, G.~Molenberghs, G.~Verbeke, and M.~Kenward.
\newblock A sensitivity analysis for shared-parameter models for incomplete
  longitudinal data.
\newblock \emph{Biometrical Journal}, 52:\penalty0 111--125, 2010.

\bibitem[Creemers et~al.(2011)Creemers, Hens, Aerts, Molenberghs, Verbeke, and
  Kenward]{cre2011}
A.~Creemers, N.~Hens, M.~Aerts, G.~Molenberghs, G.~Verbeke, and M.~Kenward.
\newblock Generalized shared-parameter models and missingness at random.
\newblock \emph{Statistical Modelling}, 11:\penalty0 279--310, 2011.

\bibitem[Dunson and Xing(2009)]{dun2009}
D.~Dunson and C.J. Xing.
\newblock Nonparametric bayes modeling of multivariate categorical data.
\newblock \emph{Journal of the American Statistical Association}, 104:\penalty0
  1042--1051, 2009.

\bibitem[Folstein et~al.(1975)Folstein, Folstein, and McHig]{fol1975}
M.~Folstein, S.~Folstein, and P.~McHig.
\newblock Mini-mental state: a pratical method for grading the cognitive state
  of patients for the clinician.
\newblock \emph{Journal of Psychiatry Research}, 12:\penalty0 189--198, 1975.

\bibitem[Gao(2004)]{gao2004}
S.~Gao.
\newblock A shared random effect parameter approach for longitudinal dementia
  data with non-ignorable missing data.
\newblock \emph{Statistics in Medicine}, 23:\penalty0 211--219, 2004.

\bibitem[Gao et~al.(2016)Gao, Hedeker, Mermelstein, and Xie]{Gao2016}
W~Gao, D.~Hedeker, R.~Mermelstein, and H.~Xie.
\newblock A scalable approach to measuring the impact of nonignorable
  nonresponse with an ema application.
\newblock \emph{Statistics in Medicine}, 35:\penalty0 5579--5602, 2016.

\bibitem[Jansen et~al.(2003)Jansen, Molenberghs, Aerts, Thijs, and
  Van~Steen]{Jans2003}
I.~Jansen, G.~Molenberghs, M.~Aerts, H.~Thijs, and K.~Van~Steen.
\newblock A local influence approach applied to binary data from a psychiatric
  study.
\newblock \emph{Biometrics}, 59:\penalty0 410--419, 2003.

\bibitem[Karlis(2001)]{karl2001}
D.~Karlis.
\newblock A cautionary note on the {EM} algorithm for finite exponential
  mixtures.
\newblock Technical Report 150, Department of Statistics, Athens University of
  Economics and Business, 2001.

\bibitem[Little(1995)]{Little1995}
R.J.A. Little.
\newblock Modeling the drop-out mechanism in repeated-measures studies.
\newblock \emph{Journal of the American Statistical Association}, 90:\penalty0
  1112--1121, 1995.

\bibitem[Little and Rubin(2002)]{litrub2002}
R.J.A. Little and D.~Rubin.
\newblock \emph{Statistical analysis with missing data, 2nd edition}.
\newblock Wiley, 2002.

\bibitem[Ma et~al.(2005)Ma, Troxel, and Heitjan]{ma2005}
G.~Ma, A.B. Troxel, and D.F. Heitjan.
\newblock An index of local sensitivity to non-ignorability in longitudinal
  modeling.
\newblock \emph{Statistics in Medicine}, 24:\penalty0 2129--2150, 2005.

\bibitem[Molenberghs et~al.(2008)Molenberghs, Beunckens, Sotto, and
  Kenward]{mol2008}
G.~Molenberghs, C.~Beunckens, C.~Sotto, and M.~G. Kenward.
\newblock Every missing not at random model has got a missing at random
  counterpart with equal fit.
\newblock \emph{Journal of the Royal Statistical Society, Series B},
  70:\penalty0 371--388, 2008.

\bibitem[Neuhaus and McCulloch(2011{\natexlab{a}})]{nehu2011a}
J.M. Neuhaus and C.E. McCulloch.
\newblock Misspecifying the shape of a random effects distribution: Why getting
  it wrong may not matter.
\newblock \emph{Statistical Science}, 26:\penalty0 388--402,
  2011{\natexlab{a}}.

\bibitem[Neuhaus and McCulloch(2011{\natexlab{b}})]{nehu2011b}
J.M. Neuhaus and C.E. McCulloch.
\newblock Prediction of random effects in linear and generalized linear models
  under model misspecification.
\newblock \emph{Biometrics}, 67:\penalty0 270--279, 2011{\natexlab{b}}.

\bibitem[Oakes(1999)]{oak1999}
D.~Oakes.
\newblock Direct calculation of the information matrix via the em algorithm.
\newblock \emph{Journal of the Royal Statistical Society, Series B},
  61:\penalty0 479--482, 1999.

\bibitem[Rakhmawati et~al.(2016)Rakhmawati, Molenberghs, Verbeke, and
  Faes]{rakh2016}
T.W. Rakhmawati, G.~Molenberghs, G.~Verbeke, and C.~Faes.
\newblock Local influence diagnostics for hierarchical count data models with
  overdispersion and excess zeros.
\newblock \emph{Biometrical journal}, 58:\penalty0 1390--1408, 2016.

\bibitem[Rakhmawati et~al.(2017)Rakhmawati, Molenberghs, Verbeke, and
  Faes]{rakh2017}
T.W. Rakhmawati, G.~Molenberghs, G.~Verbeke, and C.~Faes.
\newblock Local influence diagnostics for generalized linear mixed models with
  overdispersion.
\newblock \emph{Journal of Applied Statistics}, 44:\penalty0 620--641, 2017.

\bibitem[Rizopoulos et~al.(2008)Rizopoulos, Verbeke, Lesaffre, and
  Vanrenterghem]{riz2008}
D.~Rizopoulos, G.~Verbeke, E.~Lesaffre, and Y.~Vanrenterghem.
\newblock A two-part joint model for the analysis of survival and longitudinal
  binary data with excess zeros.
\newblock \emph{Biometrics}, 64:\penalty0 611--619, 2008.

\bibitem[Royall(1986)]{Royall1986}
Richard~M Royall.
\newblock Model robust confidence intervals using maximum likelihood
  estimators.
\newblock \emph{International Statistical Review}, 54:\penalty0 221--226, 1986.

\bibitem[Rubin(1975)]{rub1976}
D.B. Rubin.
\newblock Inference and missing data.
\newblock \emph{Biometrika}, 63:\penalty0 581--592, 1975.

\bibitem[Scharfstein et~al.(1999)Scharfstein, Rotnitzky, and Robins]{sch1999}
D.~Scharfstein, A.~Rotnitzky, and J.~Robins.
\newblock Adjusting for nonignorable drop-out using semiparametric nonresponse
  models.
\newblock \emph{Journal of the American Statistical Association}, 94:\penalty0
  1096--1120, 1999.

\bibitem[Schwarz(1978)]{Schwarz1978}
G.~Schwarz.
\newblock Estimating the dimension of a model.
\newblock \emph{The {A}nnals of {S}tatistics}, 6:\penalty0 461--464, 1978.

\bibitem[Song et~al.(2002)Song, Davidian, and Tsiatis]{son2002}
X.~Song, M.~Davidian, and A.~Tsiatis.
\newblock Semiparametric likelihood approach to joint modeling of longitudinal
  and time-to-event data.
\newblock \emph{Biometrics}, 58:\penalty0 742--753, 2002.

\bibitem[Troxel et~al.(2004)Troxel, Ma, and Heitjan]{trox2004}
A.B. Troxel, G.~Ma, and D.F. Heitjan.
\newblock An index of local sensitivity to non-ignorability.
\newblock \emph{Statistica Sinica}, 14:\penalty0 1221--1237, 2004.

\bibitem[Tsiatis and Davidian(2004)]{tsi2004}
A.~Tsiatis and M.~Davidian.
\newblock An overview of joint modeling of longitudinal and time-to-event data.
\newblock \emph{Statistica Sinica}, 14:\penalty0 793--818, 2004.

\bibitem[Tsonaka et~al.(2009)Tsonaka, Verbeke, and Lesaffre]{tso2009}
R.~Tsonaka, G.~Verbeke, and E.~Lesaffre.
\newblock A semi-parametric shared parameter model to handle nonmonotone
  nonignorable missingness.
\newblock \emph{Biometrics}, 65:\penalty0 81--87, 2009.

\bibitem[Verzilli and Carpenter(2002)]{ver2002}
C.J. Verzilli and J.R. Carpenter.
\newblock A monte carlo em algorithm for random-coefficient-based dropout
  models.
\newblock \emph{Journal of Applied Statistics}, 29:\penalty0 1011--1021, 2002.

\bibitem[Viviani et~al.(2014)Viviani, Rizopoulos, and Alf\'o]{viv2014}
S.~Viviani, D.~Rizopoulos, and M.~Alf\'o.
\newblock Local sensitivity to non-ignorability in joint models.
\newblock \emph{Statistical Modelling}, 14:\penalty0 205--228, 2014.

\bibitem[Wang and Taylor(2001)]{wan2001}
Y.~Wang and J.~Taylor.
\newblock Jointly modeling longitudinal and event time data with application to
  acquired immunodeficiency syndrome.
\newblock \emph{Journal of the American Statistical Association}, 96:\penalty0
  895--905, 2001.

\bibitem[White(1980)]{White1980}
H.~White.
\newblock A heteroskedasticity-consistent covariance matrix estimator and a
  direct test for heteroskedasticity.
\newblock \emph{Econometrica: Journal of the Econometric Society}, pages
  817--838, 1980.

\bibitem[White et~al.(2017)White, Carpenter, and Horton]{whit2017}
I.~White, J.~Carpenter, and N.J. Horton.
\newblock A mean score method for sensitivity analysis to departures from the
  missing at random assumption in randomised trials.
\newblock \emph{Statistica Sinica}, to appear, 2017.

\bibitem[Wu and Bailey(1989)]{wu1989}
M.~Wu and K.~Bailey.
\newblock Estimation and comparison of changes in the presence of informative
  right censoring: conditional linear models.
\newblock \emph{Biometrics}, 45:\penalty0 939--955, 1989.

\bibitem[Wu and Carroll(1988)]{wu1988}
M.~Wu and R.~Carroll.
\newblock Estimation and comparison of changes in the presence of informative
  right censoring by modeling the censoring process.
\newblock \emph{Biometrics}, 44:\penalty0 175--188, 1988.

\bibitem[Xie(2008)]{Xie2008}
H.~Xie.
\newblock A local sensitivity analysis approach to longitudinal non-gaussian
  data with non-ignorable dropout.
\newblock \emph{Statistics in Medicine}, 27:\penalty0 3155--3177, 2008.

\bibitem[Xie(2012)]{Xie2012}
H.~Xie.
\newblock Analyzing longitudinal clinical trial data with nonignorable
  missingness and unknown missingness reasons.
\newblock \emph{Computational Statistics and Data Analysis}, 56:\penalty0
  1287--1300, 2012.

\bibitem[Xie and Heitjan(2004)]{Xie2004}
H.~Xie and Heitjan.
\newblock Sensitivity analysis of causal inference in a clinical trial subject
  to crossover.
\newblock \emph{Clinical Trials}, 1:\penalty0 21--30, 2004.

\end{thebibliography}
\end{document}